\documentclass[a4paper,UKenglish,cleveref,autoref,thm-restate]{lipics-v2021}
\usepackage{graphicx}
\usepackage{amsmath}
\usepackage{tikz}
\usepackage[title]{appendix}
\usepackage{hyperref}
\usepackage{microtype}

\title{Performance Anomalies in Concurrent Data Structure Microbenchmarks}

\titlerunning{Performance Anomalies in Concurrent Data Structure Microbenchmarks}

\author{Rosina F. Kharal}{University of Waterloo, Canada \and  }{rkharal@uwaterloo.ca}{https://mc.uwaterloo.ca/people.html}{}

\author{Trevor Brown}{University of Waterloo, Canada \and  }{trevor.brown@uwaterloo.ca}{http://tbrown.pro}{}

\authorrunning{R.\,F. Kharal and T.\,Brown} 
\Copyright{Rosina F. Kharal and Trevor Brown} 
\acknowledgements{
We thank the reviewers for their helpful comments and suggestions.
This work was supported by: the Natural Sciences and Engineering Research Council of Canada (NSERC) Discovery Program grant: RGPIN-2019-04227, the Canada Foundation for Innovation John R. Evans Leaders Fund with equal support from the Ontario Research Fund CFI Leaders Opportunity Fund: 38512, NSERC Discovery Launch Supplement: DGECR-2019-00048, and the University of Waterloo.
}
\relatedversion{}
\relatedversiondetails{Full Version}{https://arxiv.org/abs/2208.08469}

 %TODO mandatory; please add comma-separated list of keywords

\nolinenumbers %uncomment to disable line numbering

\hideLIPIcs  %uncomment to remove references to LIPIcs series (logo, DOI, ...), e.g. when preparing a pre-final version to be uploaded to arXiv or another public repository

\EventEditors{Eshcar Hillel, Roberto Palmieri, and Etienne Rivi\`{e}re}
\EventNoEds{3}
\EventLongTitle{26th International Conference on Principles of Distributed Systems (OPODIS 2022)}
\EventShortTitle{OPODIS 2022}
\EventAcronym{OPODIS}
\EventYear{2022}
\EventDate{December 13--15, 2022}
\EventLocation{Brussels, Belgium}
\EventLogo{}
\SeriesVolume{253}
\ArticleNo{19}
\newcommand{\imgwidth}{0.33\textwidth}

\begin{document}

\maketitle

\begin{abstract} 
Recent decades have witnessed a surge in the development of concurrent data structures with an increasing interest in data structures implementing concurrent sets (CSets). Microbenchmarking tools are frequently utilized to evaluate and compare the performance differences across concurrent data structures. The underlying structure and design of the microbenchmarks themselves can play a hidden but influential role in performance results. However, the impact of microbenchmark design has not been well investigated. In this work, we illustrate instances where concurrent data structure performance results reported by a microbenchmark can vary 10-100x depending on the microbenchmark implementation details. We investigate factors leading to performance variance across three popular microbenchmarks and outline cases in which flawed microbenchmark design can lead to an inversion of performance results between two concurrent data structure implementations. We further derive a set of recommendations for best practices in the design and usage of concurrent data structure microbenchmarks and explore advanced features in the Setbench microbenchmark.
\end{abstract}

\section{Introduction} 
\label{Intoduction}
The execution efficiency of highly parallelizable data structures for concurrent access has received significant attention over the past decade.
An extensive variety of data structures have appeared, with a particular focus on data structures implementing concurrent sets (CSets)~\cite{setbench:brown,ascylib,framework,natarajan,wenmemoryrec}.
CSets have applications in many areas including distributed systems, database design, and multicore computing.
%A CSet is an abstract data type (ADT) which offers storage, access, and updating of information concurrently by multiple threads. A CSet stores keys and provides operations to search, insert, and delete keys from the set. Insert and delete operations are also called \textit{update} operations. A CSet ADT must be implemented by a concurrent data structure. There are numerous concurrent data structures that can be used to implement CSets including trees, skip-lists, and linked-lists.
A CSet is an abstract data type (ADT) which stores keys and provides three primary operations on keys: search, insert, and delete. Insert and delete operations modify the CSet and are called \textit{update} operations. There are numerous concurrent data structures that can be used to implement CSets, including trees, skip-lists, and linked-lists. 
A CSet data structure refers to the implementation of a CSet.
Microbenchmarks are commonly used to evaluate the performance of CSet data structures, essentially performing a stress test on the CSet across varying search/update workloads and thread counts. A typical microbenchmark runs an experimental loop bombarding the CSet with randomized operations performed by threads until the duration of the experiment expires. \textit{Throughput}, number of operations performed by a CSet, is a key performance metric.
%Synthetic microbenchmarks are commonly used to evaluate the performance of various CSet data structures, essentially performing a stress test on the CSet across varying search/update workloads and thread contentions. A typical microbenchmark runs an experimental loop bombarding the CSet with randomized operations performed by threads until the duration of the experiment expires. \textit{Throughput} (number of operations per second) of a CSet is a key performance metric.
%% UPDATED HERE RK %%
%in assessing performance from a microbenchmark.
Multiple platforms for microbenchmarking exist to support CSet research. The accuracy and reliability of performance results generated from microbenchmark experiments is fundamental to concurrent data structure research. Researchers must be able to assess the performance benefit vs loss of varying concurrent implementation strategies and their overall impact on performance. Microbenchmarks are also an important tool for comparative performance analysis between different implementations of CSets.
While CSet implementations have been well studied~\cite{arbel,setbench:brown, ascylib, natarajan}, the popular microbenchmarks used to evaluate them have not been scrutinized to the same degree.
Microbenchmarking idiosyncrasies exist that can significantly distort performance results. The goal of our work is to better understand the role of microbenchmark design in performance results and attempt to minimize factors present within the microbenchmark that misrepresent performance.
%% UPDATED HERE RK %%

\begin{figure}
    \centering
    %\hspace{-0.3in}
    \begin{tabular}{cc}
    (a) Initial Microbenchmark Results  & (b) Final Microbenchmark Results \\
    \rotatebox{90}{ \small \hspace{0.3in}{\textbf{ Throughput}}} 
    \includegraphics[width=\imgwidth]{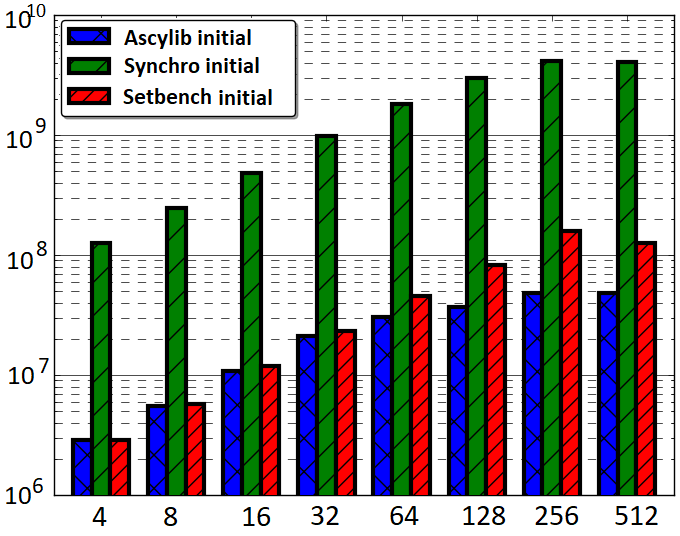} &
    \hspace{-4mm}
    \includegraphics[width=\imgwidth]{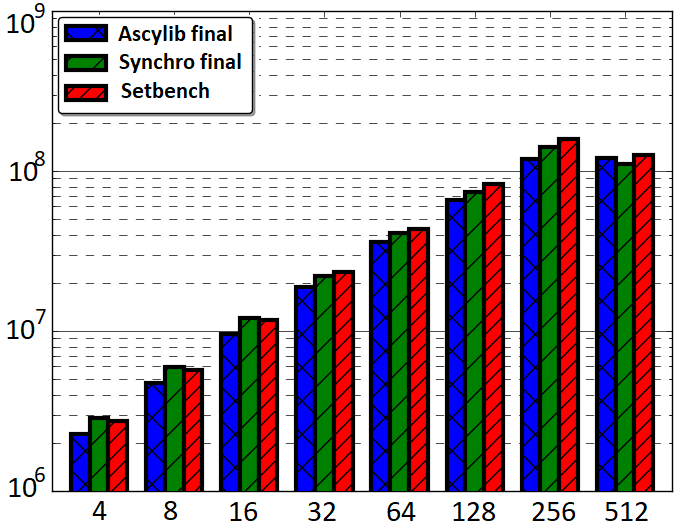} 
    \end{tabular}
    \vspace{-2.5mm}
    \caption{Throughput results across three microbenchmarks, Ascylib, Synchrobench and Setbench executing on a 256-core system testing the lock-free BST~\cite{natarajan}. Thread count is displayed on the x-axis, y-axis is a logarithmic scale. Figure (a) results from unmodified microbenchmarks (as written by their authors). Figure (b) results from modified versions of Synchrobench and Ascylib correcting for pitfalls in microbenchmark design.}
    \label{firstfig}
\end{figure}

When testing a CSet implementation on three different microbenchmarks with identical parameters, one would expect to observe similar performance results within a reasonable margin of error. However, we found 10-100x performance differences on the same CSet data structure tested across the Ascylib~\cite{ascylib}, Setbench~\cite{setbench:link}, and Synchrobench~\cite{synchro} microbenchmarks. These microbenchmarks are often employed for evaluation of high performance CSets.
In Figure 1(a) we observe a range of varying performance results on the popular lock-free BST by Natarajan et al.~\cite{natarajan} across the three microbenchmarks displayed using a logarithmic y-axis in order to capture wide performance gaps on a single scale. We performed a systematic review of the design intricacies within each microbenchmark. We found discrepancies in microbenchmark implementation leading to CSet data structures underperforming in one microbenchmark and over performing in another. We found one popular microbenchmark duplicating the entire benchmark code for each CSet implementation. This renders the code highly prone to errors related to updates or modifications to the benchmark, and may inevitably result in reporting skewed experimental results. Our investigations led to the discovery that seemingly minor differences in the architecture and experimental design of a microbenchmark can cause a 10-100x performance boost, erroneously indicating high performance of the data structure when the underlying cause is the microbenchmark itself. We performed successive updates to two of the microbenchmarks, adjusting where errors or discrepancies were discovered, until performance is approximately equalized (Figure \ref{firstfig}(b)). In this work, we discuss the primary factors leading to microbenchmark performance variance and provide a set of recommended best practices for microbenchmark experiment design. 

%%%%%%%%%%%%%%%%%%%%%%%%%%%%%%%%%%%%%%%%%%%%%%%%%%%
Microbenchmarks rely heavily on pseudo random number generators (PRNGs) to generate randomized keys and/or select randomized operations on a CSet.
In this work, in addition to investigating microbenchmark design differences, we delve into a deeper investigation of PRNG usage in microbenchmarks.
%The utilization of pseudo random number generators (PRNGs) in concurrent synthetic microbenchmarks is a heavily relied upon approach in generating randomized keys and/or deciding randomized operations on a CSet. 
Deleterious interactions between a PRNG and a microbenchmark that uses it can go undetected for years. We present examples where (mis)use of PRNGs can cause substantial performance anomalies and generate misleading results. We illustrate how using a problematic PRNG can lead to an inversion of throughput results on pairs of CSet data structures. We discuss the pitfalls of common PRNG usage in microbenchmark experiments. 
%Our experiments are limited to concurrent tree data structures that were present in the three microbenchmarks in our study.
Our experiments are limited to concurrent tree data structures that were present in the three microbenchmarks in our study. We believe the lessons learned in our investigations related to microbenchmark design apply broadly to the experimental process and are not limited to specific CSet implementations. We leave the investigation of microbenchmark performance on varying CSet implementations for future work.

\textbf{Contributions:} In this work, we perform the first rigorous comparative analysis of three widely used microbenchmarks for CSet performance evaluation. We present an overview of related work in Section \ref{relatedwork}. The three microbenchmarks evaluated in our work are described in detail in Section \ref{comparison}. We investigate the source of performance differences reported by each microbenchmark when testing equivalent tree-based data structures in Sections \ref{comparison} and \ref{memrec}. %We perform successive modifications to the benchmarks in order to equalize performance. 
We study the role of memory reclamation and its impact on CSet performance in Section \ref{memrec}. In Section \ref{randomsection}, we investigate commonly employed methods of fast random number generation and the pitfalls of each. 
%In Section 6 we investigate the role of PRNGs in microbenchmark experimental algorithms. We discuss commonly employed methods of fast random number generation and the pitfalls of each. 
%In Section 6.2 we illustrate how the (mis)use of PRNGs can lead to erroneous microbenchmark results. 
%We provide recommendations for best practice of PRNG usage in Section 6.3. 
We describe a set of derived recommendations for best practice in microbenchmark design in Sections \ref{microbenchmark_design}, \ref{APP:BST-TK} and \ref{PRNGrecommendation}. Additional recommendations for further improvements in microbenchmark design are discussed in Section \ref{other_microbenchmark_design}. Advanced features of the Setbench microbenchmark are described in Section \ref{setbench_detail} followed by concluding remarks in Section \ref{conclusions}.
In the next section, we begin with a background on the principles of microbenchmark design with concrete examples from the Setbench microbenchmark.

\section{Background}

 %we will explain how the microbenchmarks work in further detail
 %We use setbench as an example.. .we discuss synch...ascylib furhter in Section...
 %We will will look at where the perf diff comes from.
 %we will see how others differ in Sections ...
 
In this work we test three concurrent synthetic microbenchmarks, Setbench~\cite{setbench:link}, Ascylib~\cite{ascylib}, and Synchrobench~\cite{synchro} for high speed CSet analysis. The key properties of each microbenchmark are summarized in Table \ref{summary_microbenchmarks}. The microbenchmarks report the total operations per second performed on the CSet by \textit{n} threads based on a specified workload.
In particular, we study data structures that implement \textit{sets} of keys and provide operations to \textit{search}, \textit{insert} or \textit{delete} a key. 
The microbenchmarks allow users to specify parameters that include the number of threads (\texttt{t}), the experiment duration (\texttt{d}), the update rate (\texttt{u}), and the key range (\texttt{r}) contained within the set (i.e. [1, 200,000]). 
%Some microbenchmarks may also ask the user to specify the initial size of the data structure.

Evaluating performance operations on an initially empty data structure will generate results that are misleading and not representative of average performance on a non-empty CSet. Therefore, microbenchmarks typically prefill the CSet before the experiment begins to contain a subset of keys less than or equal to the total range. The prefill size may be specified by the user as the \textit{initial} (\texttt{i}) prefill amount, or the microbenchmark may decide the prefill size using its own algorithm.
For a duration \texttt{d}, a microbenchmark runs in an experiment loop where \texttt{n} threads are assigned keys from the specified key range based on a random uniform distribution, though other distributions are also possible. Threads perform a combination of search or update operations based on experimental parameters. For example, if the specified update rate (\texttt{u}) is \textit{10\%}, the search rate is \textit{90\%}. The microbenchmark either randomly splits the update rate across insert and delete operations, or employs its own algorithm to attempt to divide insert and delete operations equally. Microbenchmarks may offer the ability to specify independent insert and delete rates. This is discussed further in Section \ref{setbenchdescription} and Table 1.% across the update rate.

%%%%%%%%%%%%%%%%%%%%%% SUMMARY OF microbenchmarkS
 
\begin{table*}[tb]
\begin{small}
\hspace{-7.5mm}
    \setlength\tabcolsep{4pt}
	%\vspace{-2mm}

    %\fontsize{5pt}{10pt}
    
	%\centering
	\hspace{-11.5mm}
	\begin{tabular}{|c|c|c|c|c|c|c|}
	\hline
	\textbf{Benchmark} & Prefill Size &  Threads used & Prefill & PRNG for key & PRNG for &   \\
	\textbf{Properties} &  & to Prefill & Ops & generation & update choice &   \\
	\hline
	 Ascylib & half-full & single/n & inserts & \checkmark  & \checkmark  &  \\
	Setbench 	 & steady state$^{1}$ & n & ins/dels  & \checkmark  & \checkmark & \\
	 Synchrobench  	& half-full & single & inserts & \checkmark &  &   \\
	 
	\hline
	\textbf{Benchmark } & Centralized & Test file & Range Queries & Effective Upd & Thread & Unique  \\
	\textbf{Properties} & Test loop & per DS & Available & Option  & Pinning  & Features  \\
	\hline
	Ascylib & \checkmark   & \checkmark &  & \checkmark & \checkmark & \footnotesize *2 \\
	Setbench 	 & \checkmark & & \checkmark  &   & \checkmark & \footnotesize *3  \\
	Synchrobench  	& &  \checkmark &  & \checkmark & \checkmark  &  \footnotesize *4\\
	%\hline
	%\textbf{Experiment }			   & Centralized  & Test File  & PRNG for Key & PRNG for   &  &  \\
	%\textbf{Properties} & Test Loop & per CSet & generation & update choice & &  &
%	\hline
	%Ascylib \record\ 	& \checkmark & \checkmark & \checkmark & \checkmark &  &  \\
%	Setbench 	 & \checkmark& & \checkmark  & \checkmark  &  &  \\
	% Synchrobench  	& & \checkmark & \checkmark &  &  &  \\
	\hline
	\end{tabular}
	\vspace{2.5mm}
	\hspace{-0.4in}
	\caption{
%\trevor{timing assumptions / model of computation, whether they need htm, only support some d.s., progress}
	Summary of properties within each microbenchmark. {\footnotesize 1:} steady state depends on experiment parameters. Setbench allows varying insert and delete ratios; this determines what the data structure will fill to in steady state. {\footnotesize (*2: performance tracking. *3: statistics tracking, performance recording, automated graph generation, range query searches, varying distributions of keys/operations are possible, independent insert and delete rates possible. *4: track effective updates, alternating updates possible.)} 
	}
	\label{summary_microbenchmarks}
	\vspace{-2mm}

\end{small}
\end{table*}

\subsection {Microbenchmark Setup} 
We use the \textit{Setbench} microbenchmark as a case study to explain some underlying design principles in concurrent microbenchmarks. 
%Setbench is designed to compare the performance of concurrent set data structures that offer operations to insert, delete and search for keys.
A typical Setbench experiment involves \texttt{n} threads accessing a CSet for a fixed duration. During this time, each thread performs search or update operations that are chosen according to a specified probability distribution on keys randomly drawn from another probability distribution over a \textit{fixed} key range. 
For example, threads might choose an \textit{operation} to perform \textit{uniformly} (1/3$^{rd}$ insert, 1/3$^{rd}$ delete and 1/3$^{rd}$ search operations), and then choose a \textit{key} to insert, delete or search for from a \textit{Zipfian} distribution.
Each thread has a PRNG object, and the same object is used to select a random operation and generate random keys.

To ensure that an experiment measures performance as it would in the steady state (after the experiment has been running for a sufficiently long time), performance measurements are not taken until the data structure is warmed up by performing insertions and deletions until the CSet converges to approximately its steady state. This step is called prefilling.
%Setbench {\textit{prefill}}s the data structure using randomly generated keys from the key range, performing uniformly random insert or delete operations. %, until there is convergence to the steady state.
If the key range is [1, $10^6$], and threads do 50\% insertions and 50\% deletions, then the size of the CSet in steady state will be approximately 500,000 (half full). Different microbenchmarks will employ varying methods of prefilling the data structure prior to experimental evaluation. This is discussed in the next section. 
In this work, we evaluate performance results across three microbenchmarks and analyse the underlying subtleties in microbenchmark design which lead to varying performance on equivalent CSet data structures.
An example of this is illustrated in Figure \ref{firstfig}(a) where microbenchmark experiments are performed on the lock-free BST by Natarajan et al.~\cite{natarajan}. The initial comparative throughput results are very different. 
%We closely investigate the architectural designs of each microbenchmark in an attempt to better understand the cause of large performance gaps.% between microbenchmarks testing the same CSet data structure. 
We apply successive modifications to the microbenchmarks where required in an attempt to minimize large performance gaps. This process is outlined step-by-step in Section \ref{comparison}.

Experiments performed in this work execute on a dual socket, AMD EPYC 7662 processor with 256 logical cores and 256 GB of RAM. DRAM is equally divided across two NUMA nodes. %When running microbenchmark experiments, memory pages are interleaved across the system’s NUMA nodes (using the numactl command) to limit memory access restrictions. 
We use a scalable memory allocator, \textit{jemalloc}~\cite{jemalloc}, to prevent memory allocation bottlenecks. Each microbenchmark employs its own PRNG for generating random numbers during the experiment loop. This is discussed further in Section \ref{comparison}. We test key ranges between 2000 and 2 million keys using thread counts of up to 512, which gives an indication of the effects of oversubscribing the cores. All figures in Sections \ref{comparison} and \ref{memrec} in this work are displayed using a logarithmic y-axis in order to allow visual comparison between algorithms with large differences. 
\section{Related Work}\label{relatedwork}
There are previous efforts in the literature to better understand the underlying structure and design of benchmarks used to evaluate concurrent applications.
In their work on the comparative evaluation of transactional memory (TM) systems, Nguyen et al.~\cite{scalableTM} discuss the unexpected low performance results observed when using benchmarks to evaluate various hardware transactional memory (HTM) and software transactional memory (STM) systems. They argue that the observed limited performance results are a consequence of the programming model and data structure design used within the benchmarks and are not necessarily indicative of true performance results of the TM systems themselves. In related work by Ruan et al.~\cite{ruan}, the STAMP benchmark suite~\cite{stamp} used for evaluating transactional memory was identified as being out-of-date. The authors present several suggested modifications to the benchmark suite to boost the reliability of performance results for more accurate TM evaluation. McSherry et al.~\cite{COST} discuss the COST (Configuration that Outperforms a Single Thread) associated with scaling applications to support multi-threaded execution, and the need to measure performance gains without rewarding the substantial overhead costs of parallelization.

Recent microbenchmarks exist that were not tested in our work, such as the Synch framework~\cite{framework} for concurrent data structures evaluation. We leave this for future study. There has been some prior investigation of microbenchmark design for concurrent data structure performance evaluation. Microbenchmark experiments executing a search-only workload on CSets have been tested in previous work by Arbel et al.~\cite{arbel}. They considered differences in concurrent tree implementations of CSets and their impact on performance. It was discovered that subtle differences in concurrent tree implementations can play a pivotal role in microbenchmark performance results. 
%and discovered subtleties in varying implementations that can lead to significant changes in performance. 
Our work concentrates on the impact of \textit{microbenchmark implementation differences} on CSet data structure performance for workloads that include updates.
%We hope to perform future analysis of additional microbenchmarks such as the \textit{Synch}~\cite{framework} framework for concurrent data-structures.
%Wen et al. work with benchmarks which incorporate memory reclamation~\cite{wenmemoryrec} and perform a comparative analysis of memory reclamation algorithms. Microbenchmark memory reclamation is not our primary focus. We address microbenchmark results which incorporate memory reclamation in Section \ref{memrec}.
Mytkowicz et al. in their work, ``\textit{Producing Wrong Data Without Doing Anything Obviously
Wrong!}''~\cite{producing}, illustrate how subtle changes to an experiment’s setup can lead to enormous performance differences and ultimately to incorrect conclusions. 
Tim Harris' presentation, ``\textit{Benchmarking Concurrent Data Structures}''~\cite{tim2}, is closely related to our work. Harris explains the need for sound experimental methodology in performance evaluation tools and discusses some noted pitfalls in the Synchrobench microbenchmark in~\cite{tim}. Important considerations in the design of good concurrent data structure experiments have been previously discussed in presentations by Trevor Brown~\cite{rare}. Brown discusses subtle aspects of microbenchmark testing configurations and underlying memory and thread distributions that can play a crucial role in performance results. This is discussed further in Section \ref{other_microbenchmark_design_details}.
In our work, we provide an investigative approach to microbenchmark design by comparing design strategies employed in three popular microbenchmarks.

%%%%%%%%%%%%%%%%%%%%%%%%%%%%%
%\begin{adjustbox}
\begin{figure}
   \hspace{-11.0mm}
   % \adjustbox{width=\textwidth}{height=\textheight}{
    \begin{tabular}{ccc}
    (a) 0\% Updates & (b) 20\% Updates  & (c) 100\% Updates \\  % (c) 50\% Updates &
    \rotatebox{90}{ \small \hspace{0.3in}{\textbf{ Throughput}}} 
    %\rotatebox{90}{\textbf{10   10   10  10}}
    %\hspace{-4mm}
    \includegraphics[width=\imgwidth]{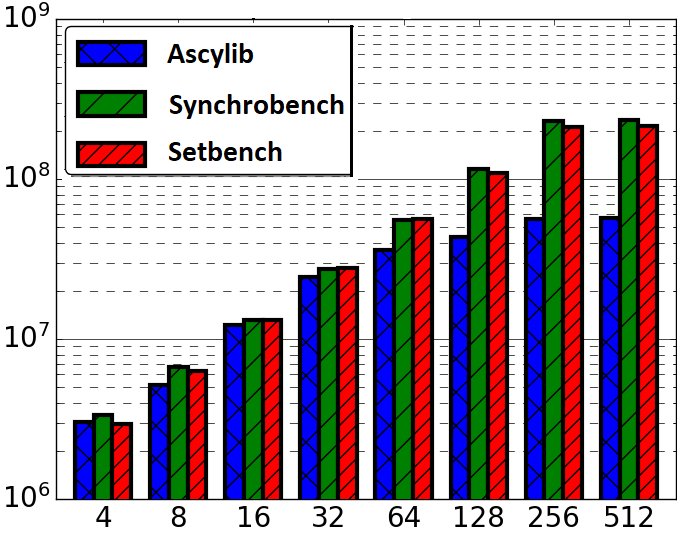} & 
    \hspace{-4mm}
    \includegraphics[width=\imgwidth]{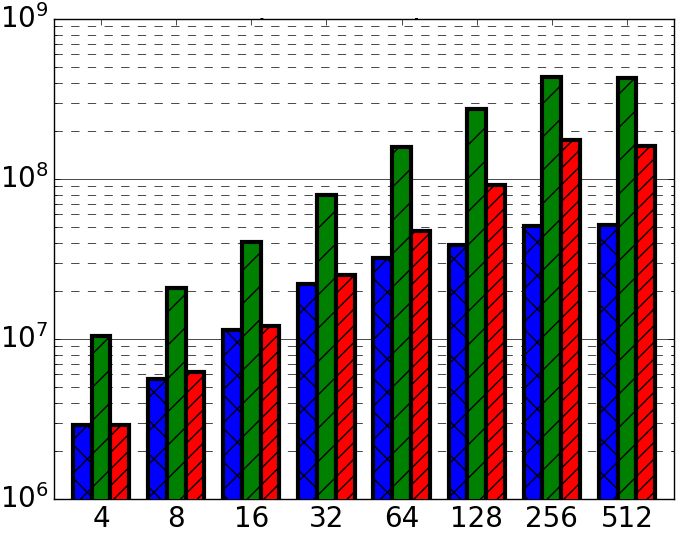} & %\includegraphics[width=\imgwidth]{figure/LFBST_Compare_Initial/NLG_50upd_Instl_Tp-2000000k_.png} &
    \hspace{-4mm}
    \includegraphics[width=\imgwidth]{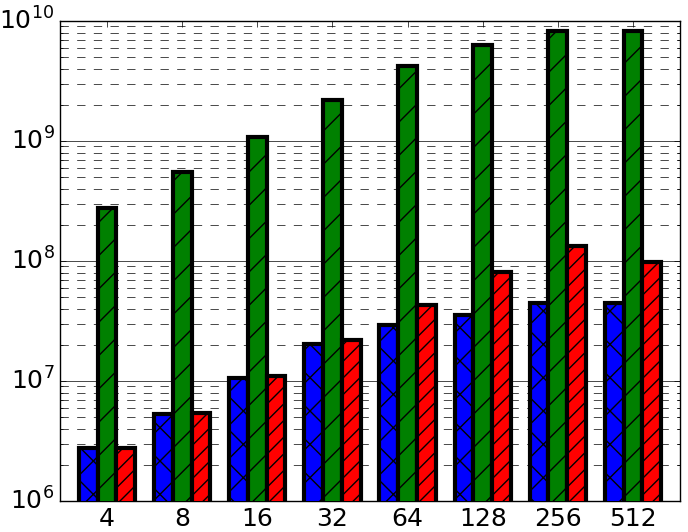} \\
    \end{tabular}
    \vspace{-2.5mm}
    \caption{Initial throughput comparisons across three unmodified microbenchmarks testing the lock-free BST~\cite{natarajan} on varying update rates.}
   
    %\belowcaptionskip{1}
    \label{install_compare}
\end{figure}
\vspace{-2mm}
%\end{adjustbox}
%%%%%%%%%%%%%%%%%%%%%%%%%%%%%%%%%%%%%%%%%%%%%%%%%%%

\section{Comparison Lock-Free Binary Search Tree} 
\label{comparison}
As mentioned above, a key performance indicator in the evaluation of CSet data structure performance is the total number of operations per second (throughput). This is computed by summing the total number of operations performed per thread and dividing by the duration of the experiment. A key indicator of memory reclamation efficiency is the \textit{maximum resident memory} occupied in RAM by the microbenchmark program during the duration of the experiment. We use this measure to evaluate the memory reclamation capabilities of each microbenchmark in Section \ref{memrec}. We perform a comparative study on the lock-free BST data structure by Natarajan et al.~\cite{natarajan} which implements a CSet. The lock-free BST stores keys in leaf nodes; internal nodes contain repeated leaf values to provide direction for searches. Not all microbenchmarks implement the lock-free BST with memory reclamation. Therefore, our initial comparisons turn memory reclamation off. Table \ref{summary_microbenchmarks} describes properties of the microbenchmarks tested in this work.
%%%%%%%%%%%%%%%%%%%%%%%%%%%%%%%%%%%%%%%%%%%%%%%%%%%%%%%%%%%%%%%%%%% 

\subsection{Synchrobench}\label{synchrodescription}
The Synchrobench synthetic microbenchmark allows the evaluation of popular C++ and Java-based CSet implementations. Synchrobench is a popular microbenchmark used for performance evaluation of CSet data structures~\cite{citesync5,ascylib,synchro, citesync6,citesync4,citesync1,citesync3}. 
Synchrobench allows users to specify an \textit{alternate option} (-A) or an \textit{effective option} (-f) as input parameters to the microbenchmark.
The -A option can be used to force threads to alternate between a key being inserted and the same key being deleted. The -f option sets total throughput calculations to count failed update operations as search operations and not as update operations. We do not use either of these options in our experiments. Synchrobench performs single threaded prefilling with insert-only operations. Each data structure directory contains a test file (\texttt{test.c}) that runs the basic test loop of the microbenchmark, performing a timed search/update workload on the CSet. The \texttt{test.c} file is repeated in each data structure directory. We discuss the potential drawbacks of this approach in Section \ref{syntestc}.
Synchrobench allows users to specify a single update rate that is divided between insert and delete operations, though the division is not necessarily equal. This is also discussed further in Section \ref{synchro_insertsvsdeletes}. 
%For example, if the update rate is 20\%, threads will perform 80\% search operations and 20\% of operations will be divided amongst insert and delete operations.

\subsection{Ascylib}\label{Ascylibdescription}

The Ascylib synthetic microbenchmark is another microbenchmark used to compare performance of concurrent data structures~\cite{arbel,ascylib,citeascy3,synchro,citeascy1,citeascy4,citeascy2}. Ascylib also performs an initial prefilling step using single threaded insert-only operations. However, Ascylib has a setting to allow multi-threaded prefilling using insert-only operations. The range and initial values are updated to the closest power of two. This is a necessary condition for the Ascylib test algorithm to generate randomly distributed keys. 
%A PRNG generates 32-bit random numbers using Marsaglia XOR-shift~\cite{marsaglia}. The random numbers are ANDed with $(range-1)$, and subsequently added to $1$ in order to ensure generated keys are within [1, range]. 
The experiment testing algorithm (\texttt{test\_simple.c}) is also repeated in each data structure directory. However, the main test loop is implemented in one common macro and is shared across each CSet data structure implemented in Ascylib. The update rate is randomly distributed among insert and delete operations and updates are not required to be effective.
Ascylib allows additional user inputs to define profiling parameters which are not tested in this work. 
%Additional properties of Ascylib can be seen in Table \ref{summary_microbenchmarks}.

\subsection{Setbench}\label{setbenchdescription}
The Setbench synthetic microbenchmark is another benchmarking tool employed in the concurrent data structure literature~\cite{arbel,setbench:link,setbench:brown,citeSB1,citeSB4,citeSB2}. Setbench also employs a directory structure for each CSet implementation. However, each CSet utilizes a single experiment test loop via an adapter class which imports each specific CSet implementation into the main experimental algorithm. This allows a single point of update for the testing algorithm and avoids software update errors. Setbench allows specification of independent insert and delete rates. Setbench uses per thread PRNGs initialized with unique seeds. Although Setbench has multiple choices of PRNGs, we employed the \textit{murmurhash3} (MM3)~\cite{murmurhash3}, a multiplicative hash function, for comparative microbenchmark experiments in this section of our work. Setbench employs multi-threaded prefilling using randomized insert and delete operations. The benefits of this are discussed in Section \ref{microbenchmark_design}. We delve into further details regarding the Setbench microbenchmark in Section \ref{setbench_detail}.

%A PRNG which was utilized by Setbench in the past led to misleading experimental results. This is further discussed in Section \ref{fnv1adescription}.
%PRNG\ref{randomsection}.
%The prefilling step in Setbench performs randomized inserts and deletes with $n$ threads until a steady state is achieved. This was described in the previous section. Using $n$ threads to perform random inserts and delete in the prefilling step allows the data structure to reach steady state naturally rather than artificially through single-thread insert-only operations. In our experiments, we noticed there is significant setup time overhead when prefilling with a single thread versus $n$ working threads.
%Setbench collects global statistical data throughout experiments and collects profiling information based on the underlying architecture. An extensive report of experimental results is output to the user including a breakdown of total operations.
%\textbf{NOTE} : Possibly MORE Setbench description

%%%%%%%%%%%%%%%%%%%%%% SUMMARY OF Versions of SYNCHRO and ASCYLIB
%\hspace{-7mm}
 %	\label{summary_synchro_ascy}
\begin{table*}[tb]
\begin{small}
    \setlength\tabcolsep{4pt}
	%\vspace{-2mm}
	\hspace{-4mm}
	\centering
	\begin{tabular}{|c|c|c|c|c|c|c|c|}
	\hline
	\textbf{Synchro Version} & Synchro  & Synchro1 & Synchro2 & Synchro3 & Synchro4 & Synchro5 & Synchro'   \\
	\hline
	 insert \& delete &  & \checkmark & \checkmark & \checkmark & \checkmark & \checkmark &       \\
	 random seeds/thread 	 & & & \checkmark  & \checkmark  & \checkmark & \checkmark &  \\
	 MM3 RNG  	& &  &  & \checkmark & \checkmark  & \checkmark &      \\
	  randomized updates  	& &      &      &      & \checkmark &  \checkmark &      \\
	  common DS impl  	& &      &      &      &      &       \checkmark & \checkmark \\

	\hline
	\textbf{Ascylib Version}  & Ascylib  & Ascylib1 & Ascylib2 & Ascylib3 & Ascylib' & &  \\
	\hline
	disable thread pin &  & \checkmark & \checkmark & \checkmark &  &   &    \\
	 
	 MM3 RNG  	& &      & \checkmark & \checkmark &   &  &   \\
	 
	  common DS impl  	& &      &      & \checkmark & \checkmark &   &   \\
%	\begin{tabular}{@{}l@{}}Can traverse pointer from\\retired \record\ to retired \record\end{tabular} &      & \xmark &      & \xmark & \xmark &      &      & \xmark &      &      &      \\
	\hline
	\end{tabular}
	\vspace{2.5mm}
	\caption{
	Summary of Synchrobench and Ascylib modifications tested in this work. Ascylib required fewer changes, Setbench did not require any modifications for the comparative experiments performed in Section \ref{comparison}. The original installed implementations are labelled Ascylib, Setbench and Synchro without a version number. Ascylib' and Synchro' are versions of each microbenchmark where \textbf{only} the lock-free BST implementation is modified (imported from Setbench).}
	\label{summary_synchro_ascy}
	%\vspace{2mm}

\end{small}

\end{table*}
\subsection{Throughput Comparisons}\label{ThroughputInstallExperiments} 

We test the initial installed implementations of the three aforementioned microbenchmarks in order to compare performance results on the lock-free BST data structure. To standardize experiments across the three microbenchmarks, we performed single threaded prefilling using insert-only operations to reach a start state where the data structure contains exactly half of the keys from the specified input range. Memory reclamation was turned off in all microbenchmarks. Attempted updates and effective updates are both counted towards the total operation throughput. We examine throughput results for experiments running for 20 seconds with update rates varying from 0\% to 100\% and a specified range of 2 million keys unless stated otherwise. Enforcing the range to a power of 2 is turned off in Ascylib experiments to match the other microbenchmarks. 
%The Ascylib random number generator algorithm is adjusted accordingly to generate random keys in the specified range. 
Initial results across the three microbenchmarks can be seen in Figure \ref{install_compare} where throughput values are displayed on a logarithmic y-axis. We observe a range of varying performance results on the lock-free BST across the three microbenchmarks. In particular, across all experiments, Synchrobench throughput results are one to two orders of magnitude higher than Setbench or Ascylib.
%The bottom row illustrates throughput results when the cores are oversubscribed. 
Ascylib results are notably lower than those of Setbench and Synchrobench. We also observe that Ascylib throughput results tend to plateau at about 128 threads and do not indicate growth as is expected and seen with Setbench and Synchrobench. We investigate further to understand the role of individual microbenchmark design on performance results.

%The Ascylib PRNG algorithm is adjusted to account for range values that are not exponential base two and to ensure the generated random keys fall between [1,range].
%%%%%%%%%%%%%%%% begin figure %%%%%%%%%%%%%%%%%%%  SYNCHRO FIXES
\begin{figure}
   \hspace{-11.0mm}
   % \adjustbox{width=\textwidth}{height=\textheight}{
    \begin{tabular}{ccc}
    (a) Synchro 50\%  & (b) Synchro 100\%  & (c) Ascylib Updates  \\
    \rotatebox{90}{ \small \hspace{0.3in}{\textbf{ \small Throughput}}} 
    \includegraphics[width=\imgwidth]{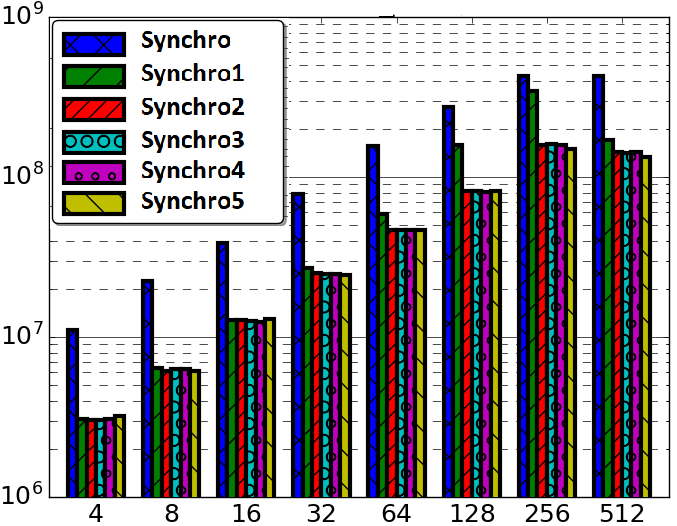} &
    \hspace{-4mm}
    \includegraphics[width=\imgwidth]{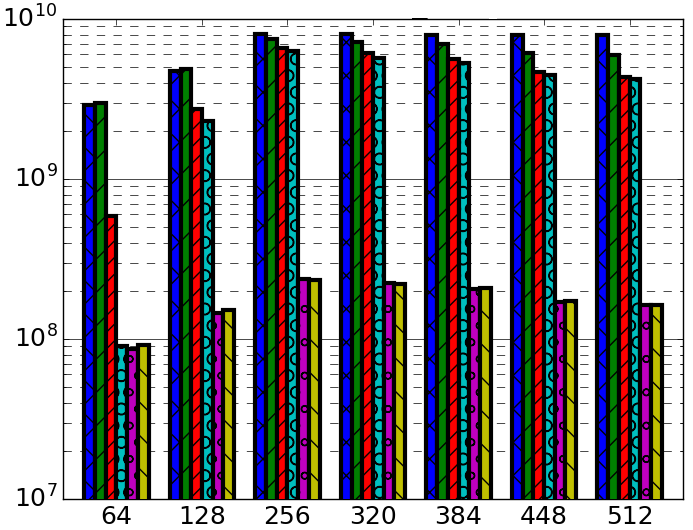} &
    \hspace{-4mm}
    \includegraphics[width=\imgwidth]{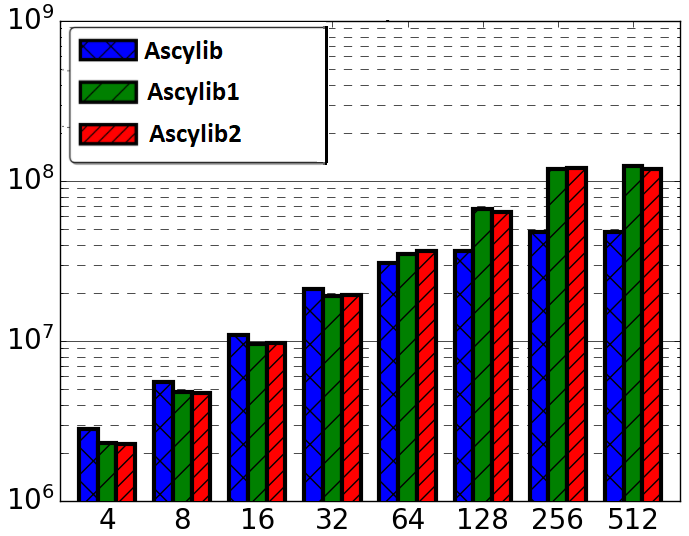} 
    \end{tabular}
    %\vspace{-2.0mm}
    \caption{Throughput results for successive modifications to Synchrobench (a), (b) and successive modifications to Ascylib (c). Synchro1 to Synchro4 involve updates to the Synchrobench microbenchmark design. Synchro5 updates the data structure implementation to that of Setbench. Figure (a) uses a key range of 2 million. Figure (b) uses a key range of 20,000 keys and displays the impact of successive modifications to Synchrobench at a 100\% update rate. 
    %Synchro is the original installed version, Synchro inserts added to experimentation, fix2 ()....Figure (c) comparison between Installed Setbench, Fix4 Synchrobench and Fix1 Ascylib at  50\% and, update 50\% search workload.
    }
    \label{syn_ascy_fixes}
\end{figure}
%%%%%%%%%%%%%%%%%%%%%%%%%%%%%%%%%%%%%%%%%%%%%%%%%%%

\subsection{Performance Factors: Synchrobench}\label{synchro_performance}
%\label{lfbstchanges_synchro}
Further investigation is required to understand the underlying causes of comparatively spiked performance results from the Synchrobench microbenchmark seen in Figure \ref{install_compare}. In the following set of experiments we aim to equalize the performance results of Setbench and Synchrobench through various adjustments made to Synchrobench where errors or bugs were discovered. We modify the original installation of Synchrobench and title each updated version as \texttt{SynchroX}, where X is the adjustment number. With each successive modification, for both Ascylib and Synchrobench, all previous modifications are maintained unless stated otherwise. A summary of modifications performed in our work are listed in Table \ref{summary_synchro_ascy}.
%%%%%%%%%%%%%%%%%%%%%%%%%%%%%%%%%%%%%%%%%%%%%%%%%%%%%%%%%%%
%\bf{Experimental Setup Bugs}
\subsubsection{Missing Insertions}
\label{syntestc}
Synchrobench utilizes a file (\texttt{test.c}) in each data structure implementation in order to run the microbenchmark experiment loop. Each thread executes in the loop for the duration of the experiment, and all threads are joined prior to termination. Insert operations occur only following a successful delete operation which indicates success by setting a variable \texttt{last} to -1. This value is checked on the next update operation; if \texttt{last} is negative an insert operation will proceed. However, the \texttt{test.c} file in the lock-free BST directory contained a bug in which \texttt{last} was an unsigned type and could never take on negative values. As a result, all experiments on the lock-free BST were performing update operations comprised of deletions and never insertions. The data structure is initially prefilled to half of the specified range, but following prefilling, insert operations never take place due to this particular bug in the experiment loop. Delete-only update operations  generate notably higher throughput results since the data structure becomes empty very quickly; essentially all operations reduce to searches as the duration of the experiment increases. Upon correction of this bug, throughput results lowered significantly. Performance results of progressive adjustments to Synchrobench are illustrated in Figure \ref{syn_ascy_fixes}. This adjustment was the first of a series of modifications made to the original version of Synchrobench for the lock-free BST and is labelled \texttt{Synchro1} in the Figure. There is a drop in throughput from the original installation of \texttt{Synchro} to \texttt{Synchro1}. We note that missing insert operations in the experiment loop was not a common occurrence in other data structure directories of Synchrobench. 
 
 %Each successive fix indicates a modification to the Synchrobench microbenchmark for the lock-free BST. In this case, fix1 indicates the correction of missing insert operations per thread in the experiment loop. As noted in Figure, allowing threads to perform both insert and delete operations results in a significant drop in throughput.

%%    (a) Compare Final 50\% & (b) Add DS Common 100\% & (c) Only DS Common 100\% \\

%%%%%%%%%%%%%%%%%%%%%%%%%%%%%%%%%%%%%%%%%%%%%%%%%%%%%%%%%%%
\begin{figure}
    %\centering
    \hspace{-11.0mm}
    \begin{tabular}{ccc}
    (a) Results Compare Final  & (b) Add DS Common & (c) Only DS Common  \\
     \rotatebox{90}{ \small \hspace{0.3in}{\textbf{ \small Throughput}}} 
    \includegraphics[width=\imgwidth]{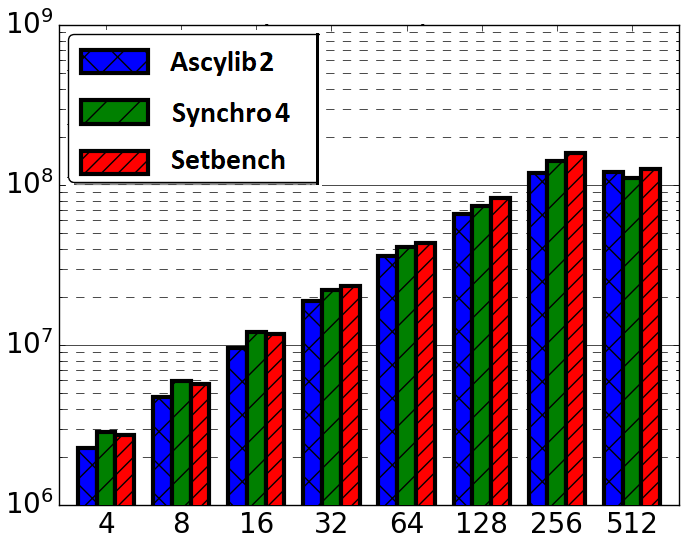} &
    \hspace{-4mm}\includegraphics[width=\imgwidth]{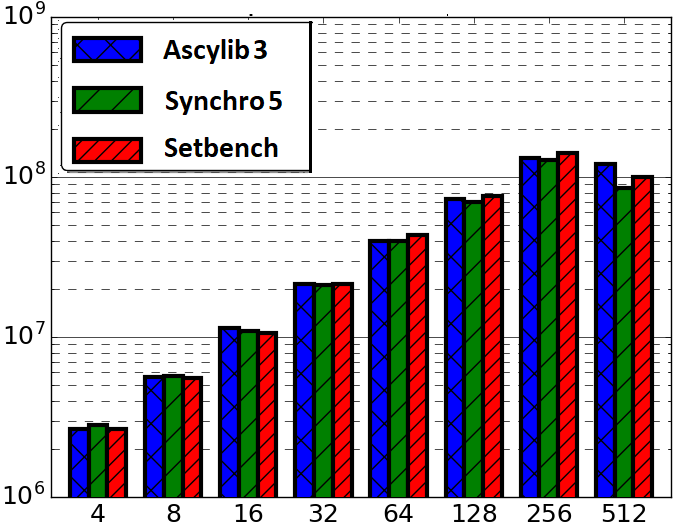} &
    \hspace{-4mm}
    \includegraphics[width=\imgwidth]{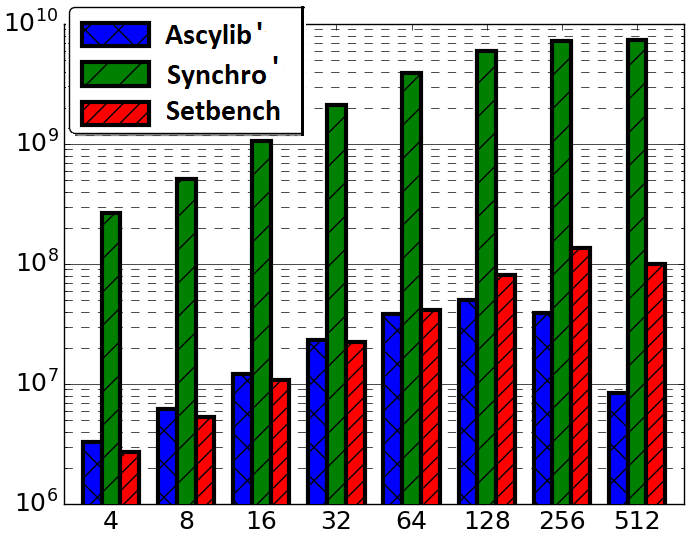} \\
    \end{tabular}
    \vspace{-2.0mm}
    \caption{ (a) Throughput results display the final comparison across three microbenchmarks with all successive modifications (Synchro4 and Ascylib2). Figure (b) tests Synchro5 and Ascylib3 which maintain all microbenchmark changes and also update the data structure (DS) implementation to that of Setbench. Figure (c) tests Ascylib' and Synchro' which do not contain any modifications to the microbenchmarks and only equalize the DS implementation.}
    \label{final_tests}
\end{figure}

\subsubsection{Thread PRNG seeds} The test algorithm (\texttt{test.c}) for the lock-free BST data structure in Synchrobench did not assign each thread a unique initialization seed for use in the PRNG employed to generate random keys. Having a PRNG initialized with the same seed per thread resulted in threads utilizing the same set of keys for search/update operations, resulting in an overall high throughput. As the duration of the experiment and the number of threads increase, updates are again essentially reduced to search operations due to other threads having previously completed the requested operation on the given key.  Inserts fail because the key is already there, deletes fail because the key was removed by another thread. With \texttt{Synchro2} we correct this problem with the addition of randomly generated seeds to initialize each thread's PRNG. The impact of this update can be seen more prominently in Figure \ref{syn_ascy_fixes}(b) where there is a drop in throughput with \texttt{Synchro2} on 100\% updates operating on a 20,000 key range. This is not so visible when the key range is much larger. At a key range of 2 million, the dominant overhead in operations is traversing a large tree; therefore, we see less variation in throughput from \texttt{Synchro2} to \texttt{Synchro5} in Figure \ref{syn_ascy_fixes}(a). The probability of contention on the same set of keys is lower at 2 million keys, therefore, the impact of \texttt{Synchro2} is more prominent in smaller key ranges.

\subsubsection{Standardized PRNG}
As discussed earlier, Synchrobench utilizes a standard built in C++ PRNG, \texttt{rand()} to supply randomly generated keys. Setbench and Ascylib use variants of XOR-shift based PRNGs. The Synchrobench microbenchmark is adapted to support the MM3 PRNG employed in Setbench. This adjustment is labeled \texttt{Synchro3} in Figure \ref{syn_ascy_fixes}. The adjustment does reduce overall throughput as MM3 uses a more complicated random number generation algorithm, using multiply and XOR-shifts, than what was previously employed in Synchrobench.

%namely murmurhash3~\cite{murmurhash3},
%HERE 
\subsubsection{Effective Insert and Delete Operations}\label{synchro_insertsvsdeletes} 
In attempt to equally distribute insert and delete operations across threads, Synchrobench uses an effective update strategy. Effective updates require threads to perform one type of update successfully before the other type of update is attempted. For example, a thread must perform and insert operation that successfully modifies the data structure before it can attempt a delete operation. This is considered an \textit{effective} update, an approach we found to offer no tangible benefit and can be unforgiving of data structure specific bugs.
Effective updates should not be confused with the -f (effective) option. The -f option in Synchrobench controls only how failed update operations will count towards total throughput, but an effective update strategy for insertions and deletions is used regardless. 

Enforcing effective updates is problematic because, for example, in an almost full data structure, to perform an effective insert, one may need to repeatedly attempt to insert many random keys until one succeeds.  Essentially, a number of search operations are inserted in between insert and delete operations, thereby inflating the total number of operations. 
The implementation of the lock-free BST in Synchrobench has a known concurrency bug contained in the original algorithm~\cite{arbel}; modified nodes are not always correctly updated in the tree. The requirement for effective updates in the experiment can generate results which erroneously indicate performance gains in the presence of errors in the implementation.
 %Enforcing an effective update before another type of update can occur offers no tangible benefit. 
%An update operation is considered \textit{effective} if it successfully modifies the CSet.  The attempts leading up to the successful insert are essentially searches (which are faster than updates), and they are counted towards throughput, thereby inflating performance.
The approach followed in Setbench is to randomize insert and delete operations using per thread PRNGs. This will generate more accurate performance results in spite of possible errors in the implementation. This adjustment is added to Synchrobench and is labelled \texttt{Synchro4}.

It may also be noted that a \textit{checksum validation} step would prove beneficial in Synchrobench to catch data structure related concurrency bugs. A checksum validation verifies that the sum of keys inserted minus the sum of keys deleted into the CSet \textit{during} an experiment should equal the final sum of keys contained in the CSet \textit{following} the experiment. Incidentally, the implementation of the lock-free BST in Synchrobench was failing checksum validation. \texttt{Synchro4} is the final correction to the Synchrobench microbenchmark design. The data structure specific concurrency bug is updated in the next modification. 
%A data structure implementation update is not considered to be an update to the Synchrobench microbenchmark design.

% certainly recommend a checksum validation in microbenchmark experiments to catch concurrency related bugs: the sum of keys inserted minus the sum of keys deleted into the CSet \textit{during} an experiment should equal the final sum of keys contained in the CSet \textit{following} the experiment.
%Incidentally, the implementation of the lock-free BST in Synchrobench was failing checksum validation. \texttt{Synchro4} is the final correction to the Synchrobench microbenchmark design. The data structure concurrency bug is updated in the next modification. 
%A data structure implementation update is not considered to be an update to the Synchrobench microbenchmark design.

\subsubsection{Equalizing the Lock-Free BST Implementation}\label{synchro5} 
The final update to Synchrobench is a modification of the data structure implementation and equalizing the three microbenchmarks to use the lock-free BST implementation provided in Setbench. The Setbench implementation corrects the concurrency bug and adds checksum validation, which does not exist in the other microbenchmarks. This adjustment is labeled \texttt{Synchro5}. We do not see a large difference in performance from \texttt{Synchro4} to \texttt{Synchro5} in Figure 3(b), which highlights the need for randomized insert and delete operations in concurrent microbenchmark experiments. By employing a randomized update operation assignment, we mitigate the impact of concurrency bugs on overall CSet data structure performance. We also assess an implementation of Synchrobench, \texttt{Synchro'}(Synchro prime), with the imported lock-free BST implementation from Setbench which \textit{does not} include any modifications to the Synchrobench microbenchmark given in \texttt{Synchro1} to \texttt{Synchro4}. This comparison is given in Figure \ref{final_tests}.
%The full set of modifications to Synchrobench and Ascylib are listed in Figure \ref{summary_synchro}.
%%%%%%%%%%%%%%%%%%%%%%%%%%%%%%%%%%%%%%%%%%%%%%%%%
%%%%%%%%%%%%%%%%%%%%%%%%%%%%%%%%%%%%%%%%%%%%%%%%%%%%%%%%%%%
\subsection{Performance Factors: Ascylib} \label{lfbstchanges_ascylib}
The Ascylib microbenchmark test algorithm and underlying default settings lead to a few factors that impact performance results on the lock-free BST. Each successive modification to Ascylib is labeled \texttt{AscylibX}.

\subsubsection{Thread Pinning} \label{Ascy_threads}
The Ascylib general installation enables thread pinning by default. With further investigation, we found that built-in thread pinning settings were under utilizing the 256 available cores during experimentation. Ascylib captures the underlying core and NUMA node count at compile time; we updated build settings to ensure the correct number of cores were detected. Although the Ascylib build displays that the correct number of cores have been detected, we found the Ascylib throughput results in Figure \ref{install_compare} were based on under 50\% core utilization. The default settings were unable to utilize the full set of cores. To remove the underlying thread pinning settings, and disable thread pinning entirely, we recompiled with SET\_CPU=0. This adjustment is labelled \texttt{Ascylib1}. Results for \texttt{Ascylib1} indicate full core utilization and improve performance in Figure \ref{syn_ascy_fixes}(c). A user that is unaware of Ascylib's default setting may unknowingly generate misleading results.
Rather than modifying the three benchmarks to perform identical thread pinning, we disabled thread pinning in all three for consistency. This is perhaps not ideal for microbenchmark experiments. Recommendations for thread pinning in microbenchmark experiments are discussed further in Section \ref{other_microbenchmark_design}.
%**UPDATED HERE RK ** \\

%For the purposes of experiments in this section we have disabled thread pinning for consistency in all microbenchmarks. The relevance of thread pinning in microbenchmark experiment design is further discussed in Section \ref{microbenchmark_design}.

%Explain why disabling thread pinning is the best approach. In fact, microbenchmakr experiments should have thread pinning enabled.... discussed in the NEW SECTION (microbenchmark design) Section 4.8

%Disabling thread pinning is consistent with Setbench and Synchrobench settings. 
%-DCORE\_NUM=256 output during compile

\subsubsection{Standardized PRNG} As was the case with the Synchrobench microbenchmark, we use the same PRNG across all three microbenchmarks. Ascylib is also updated to use the MM3 PRNG employed in Setbench. The update is labelled \texttt{Ascylib2}. We do not see a significant observable change in performance between \texttt{Ascylib1} and \texttt{Ascylib2} on a logarithmic scale. The MM3 algorithm is a more complicated PRNG (multiply, XOR-shifts) than what was previously used in Ascylib (Marsaglia XOR-shift~\cite{marsaglia}). Additional testing reveals a slight drop in performance on a non-logarithmic scale when switching the PRNG to MM3.
%which is a more complicated PRNG algorithm. The original PRNG algorithm used in Ascylib was also based on XOR-shifts~\cite{marsaglia}. 
%This is the final modification related to the Ascylib microbenchmark design. 
%The next modification is an update to the lock-free BST data structure implementation. 

 %Ascylib's installed RNG is the Marsalgia XOR-shift based RNG~\cite{marsaglia}. 

\subsubsection{Equalizing the Lock-Free BST Implementation} Last, for a comparison that evaluates a standard data structure implementation on each microbenchmark, we implement the lock-free BST implementation from Setbench into Ascylib. This is labelled \texttt{Ascylib3}. \texttt{Ascylib3} maintains all previous benchmark adjustments whereas \texttt{Ascylib'} only updates the common data structure implementation from Setbench into the original installation of Ascylib (Table \ref{summary_synchro_ascy}).

 %%\ref{summary_synchro_ascy}

\subsection{Final Comparisons}
The final comparative results following successive modifications to Ascylib and Synchrobench are given in Figure \ref{final_tests}(a), which tests \texttt{Ascylib2}, \texttt{Synchro4} and the original Setbench implementation. These implementations use the built in data structures of each microbenchmark while adjusting for microbenchmark design differences in an attempt to equalize the throughput results. We have achieved throughput results that are fairly consistent across microbenchmarks. There are slight discrepancies in throughput results, but these are not nearly as drastic as the performance differences across the original implementations of Ascylib and Synchrobench in Figure \ref{install_compare}. 
Additional final comparisons are provided in Figure \ref{final_tests}(b) and (c),
which also equalize the lock-free BST implementation across all microbenchmarks on a 100\% update workload across 2 million keys. Figure \ref{final_tests}(b) includes all microbenchmark modifications for both Synchrobench and Ascylib, whereas Figure \ref{final_tests}(c) does not include any microbenchmark modifications from the original installed versions of Synchrobench and Ascylib. 
The results in Figure \ref{final_tests}(c) illustrate the variations in throughput that occur on account of microbenchmark implementation differences. We see that once microbenchmark idiosyncrasies have been ironed out in (a) and (b), the performance results are much more consistent.
This highlights again the crucial role of microbenchmark design in the observed performance of CSet data structures.

\vspace{-2.0mm}
\subsection{Microbenchmark Design Considerations} \label{microbenchmark_design}
In this section we investigated microbenchmark idiosyncrasies between three microbenchmarks. We performed successive modifications to two of the microbenchmarks to account for design differences. During our experiments, we discovered the following factors in microbenchmark design which lead to the greatest impact on performance: (1) Repeated benchmark code is prone to error. In Synchrobench where the algorithm running performance experiments is duplicated for each data structure, errors in the algorithm led Synchrobench results to exceed other microbenchmarks by 100x. The microbenchmark testing algorithm should exist in one centralized location and provide easy adaptation to new data structures. (2) Microbenchmarks use a variety of techniques for splitting the update rate between insert and delete operations. Recommended practice is to randomly distribute update operations between inserts and deletes using per thread PRNGs. (3) Synchrobench introduced a setting to enforce effective updates. We note in Section \ref{synchro_insertsvsdeletes}, effective updates unnecessarily inflate throughput results and are not recommended. (4) Our recommended best practice for microbenchmark design includes strategies to detect and mitigate errors in the microbenchmark. We certainly recommend a checksum validation in microbenchmark experiments. In our work, adding checksum validation assisted in discovering microbenchmark and data structure implementation errors. 

Prefilling a CSet prior to running the microbenchmark experiment is also an important design consideration. Although experiments in this section used insert-only prefilling, we recommend against this for CSet microbenchmark experiments. (5) Data structure prefilling should occur through (a) randomized insert and delete operations, and (b) using the same \textbf{n} threads that will be employed during the measured portion of experiments. This will generate a more realistic configuration of a concurrent data structure in steady state as opposed to a data structure prefilled using single-threaded insert-only operations. Single-threaded prefilling will result in memory allocation specific to one thread's NUMA node. This will results in memory access latency for threads on different NUMA nodes during the measured portion of experiments. Using \textbf{n} threads to perform prefilling will disperse memory allocation across additional NUMA nodes. \textbf{N}-threaded prefilling with randomized insert and delete operations is used in Setbench as mentioned previously. We discuss additional considerations in microbenchmark design and provide further recommendations in Section \ref{other_microbenchmark_design}. In the next section, we experiment with memory reclamation in microbenchmarks and evaluate its impact on performance.

%%%%%%%%%%%%%%%%%%% TEST
\begin{figure}
    \centering
    %\hspace{-11.5mm}
    \begin{tabular}{cc}
     \small (a) Max Resident & \small (b) Throughput \\
    \hspace{-3mm}
    \includegraphics[width=0.35\linewidth]{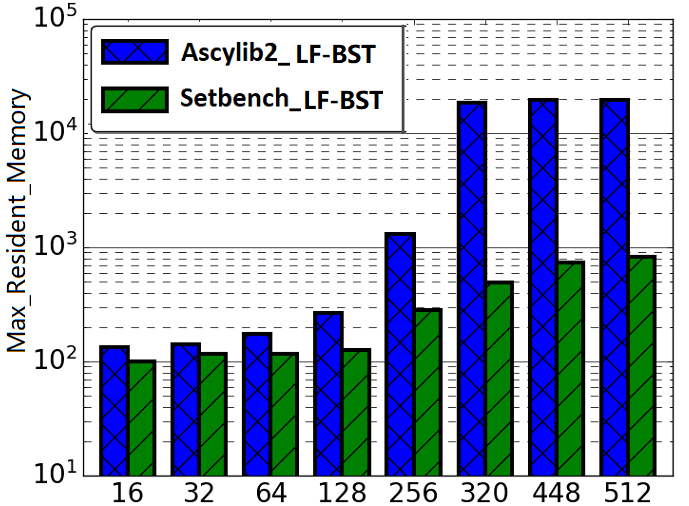} 
    %\rotatebox{90}{\hspace{0.3in} \small \textbf{Throughput}}
    \hspace{-1mm}
    & \includegraphics[width=0.35\linewidth]{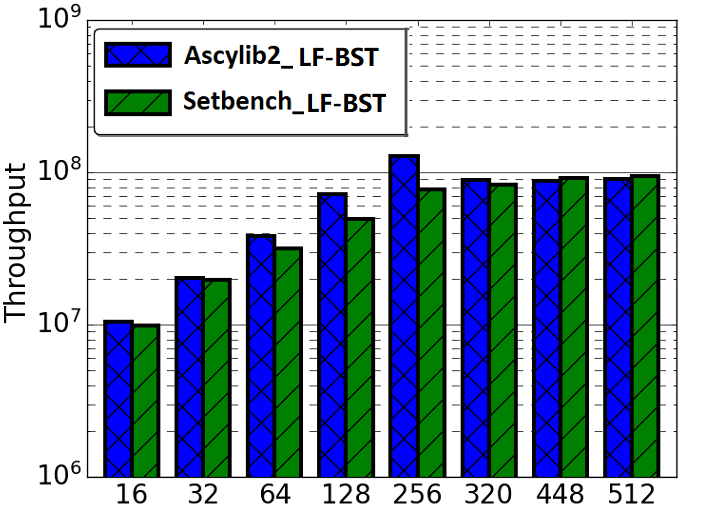} \\
    
     (c) Max Resident & \small(d) Throughput  \\

    \hspace{-5mm}
    \includegraphics[width=0.35\linewidth]{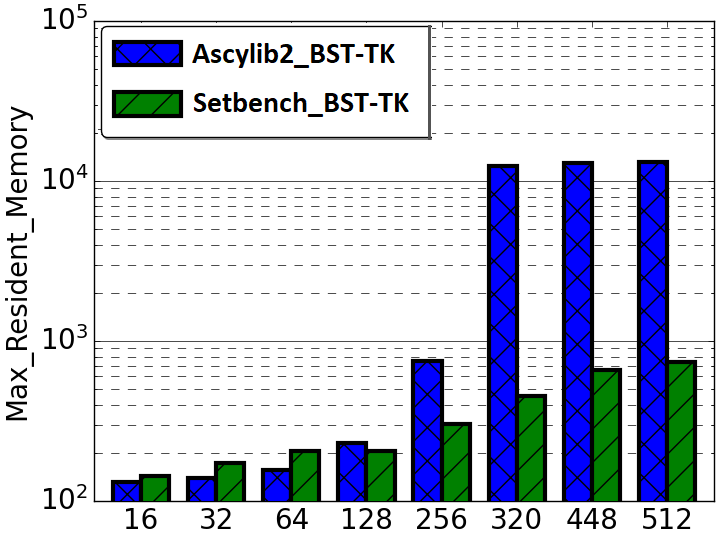}  & 
    \hspace{-2mm}
    \includegraphics[width=0.35\linewidth]{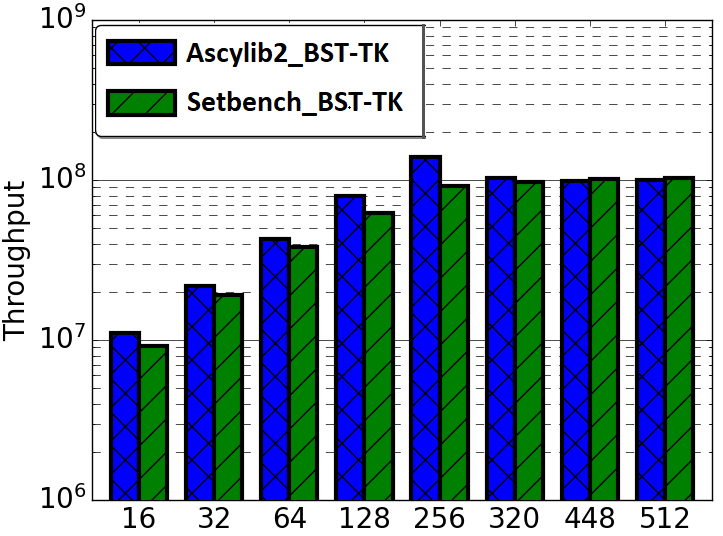} \\
    \end{tabular}
   \caption{
    Maximum Resident Memory and Throughput results for Ascylib and Setbench on the lock-free BST ((a), (b)) and BST-TK ((c), (d)). Ascylib implementation contains microbenchmark updates contained in Ascylib2 (Section \ref{Ascylibdescription}). }
    %the BST-TK data structure executing in Ascylib and Setbench on a 2 million key range at 50\% update rate. BST-TK implementation in }
   \label{bst_results}
\end{figure}

\section{Memory Reclamation}\label{memrec}

A key measure of memory usage for an executing program is the \textit{maximum resident memory} occupied in RAM by the program during the duration of its execution. 
%We examine the max resident values across the three microbenchmarks on the Natarajan lock-free BST ~\cite{natarajan}.
The lock-free BST as described by Natarajan et al.~\cite{natarajan} does not provide a complete algorithm for memory reclamation during execution. The partial algorithm suggests removing an unbounded number of nodes that are nearby neighbours in the tree pending deletion. Any given thread may proceed to delete and free (unlink) \texttt{n} nodes that are in close proximity within the tree. However, the original implementation was leaking memory. Synchrobench does not implement any memory reclamation in its implementation, whereas Ascylib has an added option for garbage collection (GC). 
The authors of the lock-free BST suggest adding epoch based memory reclamation, but it was not so simple. The memory reclamation algorithm from the original work is updated in the Setbench implementation to correctly reclaim memory~\cite{arbel}. 
%\subsection{Lock-Free BST}
We first compare the memory reclamation implementations in Setbench and Ascylib by setting Ascylib's GC setting to true, and Setbench epoch based reclamation is turned on.  We show comparative analysis of results across each microbenchmark in Figure \ref{bst_results}(a) and (b). The Ascylib microbenchmark has been updated to \texttt{Ascylib2} in order to disable thread pinning and equalize the PRNG utilized in both microbenchmarks. We have ensured all 256 cores are being utilized by Ascylib.
Figure \ref{bst_results}(a) illustrates differences in each microbenchmark's ability to reclaim memory as the thread count increases and cores are oversubscribed. Ascylib's memory usage surpasses that of Setbench by over one order of magnitude, particularly as the thread count increases. Throughput results (Figure \ref{bst_results}(b)) are relatively equal, however, the high maximum resident memory values may render Ascylib experiments unfeasible in some settings. 
We further consider microbenchmark comparisons on the equalized lock-free BST implementation with memory reclamation turned on. We discover similar performance discrepancies to those discussed in Section \ref{comparison}, although the data structure implementation and memory reclamation algorithms are identical across the three microbenchmarks. Performance results continue to show variance until microbenchmark idiosyncrasies are accounted for. Results for these additional experiments can be seen in Appendix \ref{appendixA}.

%%%%%%%%%%%%%%%%%%%%%%%%%%%%%%%%%%%%%%%%%%%%%%%%%%%

\subsection{Setbench/Ascylib BST Ticket} \label{APP:BST-TK}
In Section \ref{comparison} of this work, we examined performance factors for the lock-free BST on three concurrent synthetic microbenchmarks. We noted substantial impacts on performance as a result of microbenchmark implementation intricacies. In this section we investigate performance differences across the BST ticket (BST-TK) CSet data structure as implemented in the Setbench and Ascylib microbenchmarks. The ticket based binary search tree by Guerraoui et al.~\cite{ascylib} appears in both Setbench and Ascylib microbenchmarks; however, it is not implemented in Synchrobench.
The BST-TK is an external binary tree where leaf nodes contain the set of keys contained within the data structure. Internal nodes are used for routing and contain locks and a version number. This allows optimistic searches on the tree where concurrency can be verified by the correct version number. Both Ascylib and Setbench implement the BST-TK with memory reclamation. Ascylib has garbage collection (GC) turned on, Setbench performs epoch based reclamation. 
%We perform comparative analysis of results across each microbenchmark in Figure \ref{bst_results}. The Ascylib microbenchmark has been updated to \texttt{Ascylib2} in order to disable thread pinning and equalize the PRNG utilized in both microbenchmarks. We have ensured all 256 cores are being utilized by Ascylib.
We observe in Figure \ref{bst_results}(d) that throughput results from both microbenchmarks are similar on the BST-TK data structure. In Figure \ref{bst_results}(c), we see again, the Ascylib microbenchmark has higher memory usage; a greater than one order of magnitude increase over Setbench. This may render Ascylib experiments impractical in some settings and indicates memory is leaking at higher thread counts. 

We have seen microbenchmarks can vary greatly in performance and memory usage across two different concurrent data structure implementations. We recommend microbenchmark users investigate overall memory usage in parallel with throughput results in order to get a clear understanding of the role of memory reclamation on the performance of CSets. Memory may not be leaking necessarily; if the memory reclamation algorithm is simply slow or inefficient, there maybe a tangible impediment on performance.

%The underlying microbenchmark design and built in memory reclamation algorithm can play a significant role in overall results. 
%Memory reclamation can play an influential role in performance results. 
%The Ascylib microbenchmark memory reclamation algorithm is leaking memory at higher thread counts.
%%%%%%%%%%%%%%%% begin figure %%%%%%%%%%%%%%%%%%%
\begin{figure}
    \centering
    \includegraphics[width=\linewidth]{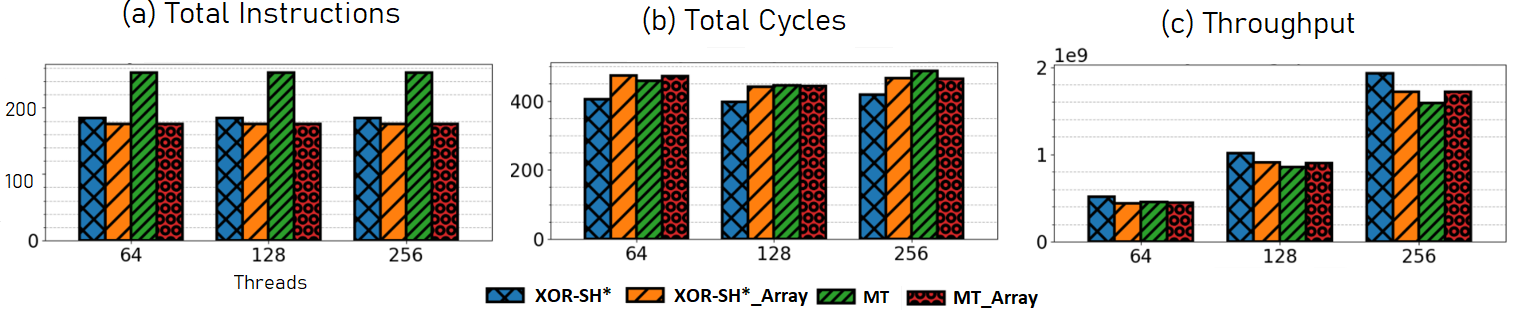}
    %\vspace{-2.0mm}
    \caption{Array Based Pre-generated PRNG vs Non Array based PRNG on a key range of 20 000. (a) Total Instructions per operation (b) Total Cycles per operation (c) Total Throughput per second}
    \label{array_results}
\end{figure}
\section{Randomness in Concurrent Microbenchmark Experiments}
\label{randomsection}
As we have seen in previous sections, concurrent microbenchmarks rely heavily on PRNGs to generate randomized keys and randomized operations for high performance CSets that can perform potentially billions of operations per second. A fast PRNG is key. In this section we draw our attention to best practices of PRNG usage in concurrent microbenchmark experiments. We limit our attention to non-cryptographic PRNGs due to the speed requirement. It is desirable to utilize a PRNG with low overhead costs to the running experiment. Some microbenchmarks may choose a custom built PRNG, while others may opt for a standard built-in PRNG such as \texttt{rand()} used in Synchrobench. Some will pregenerate an array of random numbers (RNs); this allows fast, direct access to a list of RNs and avoids in-place generation costs. If properties of high quality randomness are desired, one may use an architecture specific hardware RNG. We explore the practicality and benefit of these approaches in subsequent sections.
The PRNGs tested in this work include commonly used software PRNGs: murmurhash2 (MM2)~\cite{murmurhash2}, murmurhash3 (MM3)~\cite{murmurhash3}, Mersenne Twister (MT)~\cite{mersennetwister}, MRG~\cite{mrgprng}, and an implementation of the Marsaglia XOR-shift PRNG (XOR-SH*)~\cite{marsaglia_wiki,marsaglia}.
We describe custom hash functions and hardware RNGs in subsequent sections.
%The PRNGs tested in this work are given in Figure \ref{prnglist}.
%% I THINK FNV1a Cutom Hash function FNV1a~\cite{fnv1a:links}, 
%Murmurhash2 (MM2),
Experiments in this section were run for durations of 3-5 seconds.

\vspace{-2.0mm}
\subsection{Pre-Generated Array of Random Numbers} \label{pregenarray}
%One might be tempted to think that the best way to obtain fast, high quality randomness would be to pre-generate a large array of RNs (or one array per thread) before running an experiment. Then, one could employ hardware randomness, or a cryptographic hash function and push the high cost of generating RNs into the unmeasured setup phase of the experiment. We tested this method in Setbench with per thread arrays of pre-generated RNs. We found this approach to be counter productive due to the high cost of processor cache misses when accessing a large array. Experiments for this section are further explained in Appendix \ref{pregenarrays}.
One might be tempted to think that the best way to obtain fast, high quality randomness would be to pre-generate a large array of RNs (or one array per thread) before running an experiment. Then, one could employ hardware randomness, or a cryptographic hash function, and push the high cost of generating random numbers into the unmeasured setup phase of the experiment. We tested this method in Setbench with per thread arrays of pre-generated RNs using the XOR-SH* and MT PRNGs versus in-place RN generation with each algorithm. It is important to note that the pre-generated array approach eliminates the cost of in-place random number generation; during an experiment it is simply a matter of requesting an index into an array to generate the next random.
A limitation of an array-based approach is, of course, the array size. It is undesirable to have frequent repetition of RNs during experimentation. We use array sizes of 10 million to generate a large set of RNs. Results in Figure \ref{array_results}(c) indicate the XOR-SH*\_Array employed during experiments was notably slower than using the XOR-SH* algorithm in-place. This is due to the fact that accessing a large array of 10 million will lead to additional clock cycles generated by cache misses. An algorithm that is relatively fast, such as the XOR-SH* PRNG, will not benefit from taking a pre-generated array-based approach. However, a slightly more complex algorithm such as MT, which requires more instructions (Figure \ref{array_results}(a)), can benefit from an array-based approach. The MT\_Array generates slightly higher throughput results than using MT alone as indicated in Figure \ref{array_results}(c). However, the benefit is not as striking as one may expect with an array-based PRNG approach.
There may be use cases for an array-based PRNG such as requiring a more complicated (exotic) distribution of RNs. In this case, pre-generating RNs in an array may be an effective approach to limiting the overhead of a complex algorithm.
%%%%%%%%%%%%%%%% begin figure %%%%%%%%%%%%%%%%%%%
\begin{figure}
    \centering
    \includegraphics[width=\linewidth]{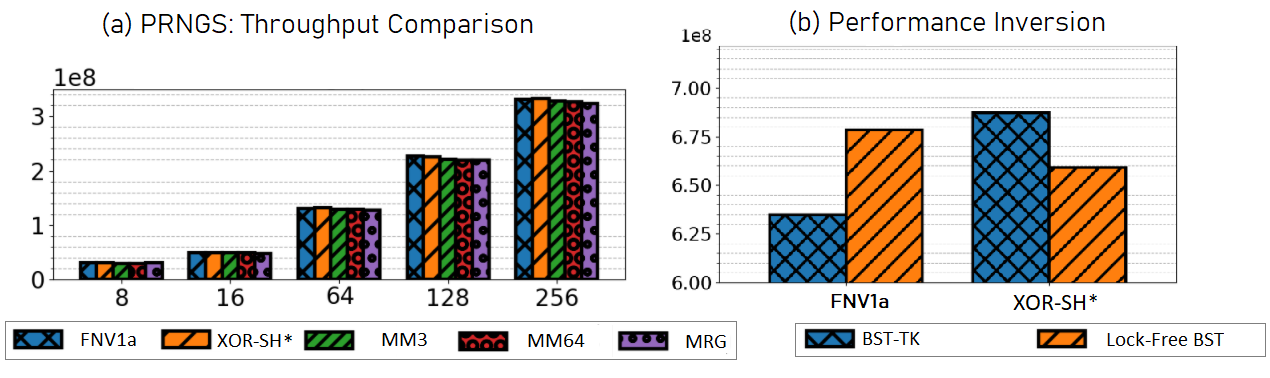}
    \caption{(a) Throughput results comparing FNV1a to other PRNGs; FNV1a algorithm does not indicate any detectable performance anomalies in throughput. Figure (b) on a smaller key range (4096), illustrates the subtle effects of a PRNG. There exists a performance inversion when using the FNV1a PRNG vs XOR-SH*.}
    \label{APP:FNV1a inversion}
\end{figure}
%%% (b) FNV1a performance inversion: 50\% inserts, 0.5\% delete workload
%%%%%%%%%%%%%%%%%%%%%%%%%%%%%%%%%%%%%%%%%%%%%%%%%%
\subsection{PRNG Associated Experimental Anomalies}
\label{fnv1a_anomaly}
In the search for high-speed generation of RNs, researchers may choose to implement their own PRNGs or use a custom hash function that may not have been well tested for properties consistent with high-quality PRNGs.
Prior to this work, the PRNG used in Setbench was FNV1a~\cite{fnv1arecommendation}, a fast, non-cryptographic 64-bit hash function. FNV1a is recommended by Lessley et al. as a hash function with \textit{``consistently good performance over all configurations''}~\cite{fnv1arecommendation}.
Setbench employed an FNV1a based PRNG that was used to generate both random operations and random keys.
However, upon testing single threaded experiments, we noticed that the data structure prefilling step was failing to converge (i.e., it was non-terminating).
%Setbench will continue to prefill not begin  until prefilling the CSet converges to the steady state.
Upon further investigation, we found that the FNV1a based PRNG was generating RNs that followed a strict odd-even pattern. That is, after generating an even number, the next number would always be odd, and vice versa. (The initial seed determined
whether the first number was even or odd.) During prefilling, a thread uses the first RN to determine the key and the second RN to determine the operation. In this case, the set of keys were always either all odd or all even, leading to an infinite loop when attempting to prefill the data structure to half full. Recall that Setbench employs both insertions and deletions to prefill the CSet to steady state (half-full in this case).
This odd-even pattern can easily be missed in overall data structure performance results. Figure \ref{APP:FNV1a inversion}(a) illustrates throughput results comparing various PRNGs tested in Setbench. There is no notable indication of threads generating all even or all odd keys from the FNV1a algorithm. Some threads are generating all even keys, while others are generating all odd keys. Setbench prefilling occurs with \textit{n} threads, so as soon as the thread count increases from 1, the probability of convergence increases.
One could imagine this kind of error remaining undetected and having a subtle effect on performance; limiting the set of keys per thread will affect which other threads it could contend with. In addition, it is not sound experimental methodology for a microbenchmark to generate keys based on this pattern.
Second, this undesirable behaviour found in FNV1a can lead to performance inversions when evaluating CSets in a microbenchmark.
The impact of the FNV1a based PRNG is more clearly displayed in the results of Figure \ref{APP:FNV1a inversion}(b) where, given a high insert workload, FNV1a can lead to a performance inversion of experimental results.
The experiment illustrates that the lock-free BST (Natarajan et al.)~\cite{natarajan} throughput results are 1.12 times higher than that of the BST-TK (Guerraoui et al.)~\cite{ascylib} when Setbench is using FNV1a as its PRNG. However, using another PRNG, such as XOR-SH*, we see the results indicate the lock-free BST \textit{underperforms} by a factor of 0.96. This is an approximately 16\% performance error leading to an inversion of results that could possibly remain undetected when one concurrent microbenchmark employs a problematic PRNG algorithm such as FNV1a. We implement a tool, the \textit{${n^{th}}-bit$ summation result}, to assess bitwise randomness in RNs generated by a PRNG. Incidentally, FNV1a also illustrated periodic behaviour in higher order bits. The tool is further discussed in Appendix \ref{appendixB}.

%In that graph, each thread in the FNV1a algorithm is generating either \textit{all even keys} or \textit{all odd keys}.
%On the other hand, in the other algorithms, each thread can generate \textit{any key}.
%There is no obvious way to tell that this is occurring by looking at such a graph where throughput results are all very similar. % that in the FNV1a algorithm, each thread is generating either \textit{all even} keys or \textit{all odd} keys.

\begin{figure}
%\hspace{-7mm}
    \centering
    \includegraphics[width=\linewidth]{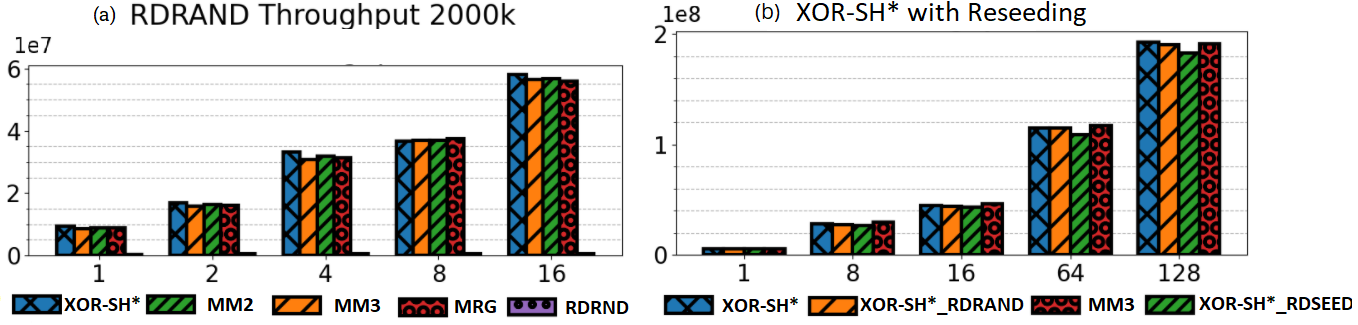}
    \caption{Hardware vs Software PRNGs: Figure (a) RDRAND throughput compared to software PRNGs. RDRAND has the lowest throughput values. In Figure (b) Reseeding XOR-SH* with RDSEED (or RDRAND) every 1 million random numbers indicates no strong penalty for reseeding a software PRNG with a hardware RNG.}
    \label{RDRAND}
\end{figure}

\subsection{Hardware RNG} \label{hardware}
A search for a high-quality 64-bit PRNG with its own source of entropy led us to an Intel Secure Key instruction, RDRAND, available on Ivy Bridge processors~\cite{Rdrand:intel}. The RDRAND instruction returns an RN from Intel on-chip RNG hardware. We compare RDRAND with the software based PRNGs to evaluate the suitability of a hardware based PRNG for use in synthetic microbenchmarks. We have illustrated throughput results in experiments with various PRNGs in Figure \ref{RDRAND}(a). We can see that the experimental throughput of RDRAND is significantly less compared to software PRNGs. A smaller key range of 2000 keys was necessary to visually illustrate the low throughput results generated when RDRAND is employed in Setbench. The overhead costs of hardware entropy greatly impede the overall performance of an experiment which aims to maximize throughput results. Algorithms such as XOR-SH* and MM3 that are computationally fast in nature have much higher throughput results. For concurrent microbenchmark experiments, it is not recommended to use a hardware PRNG alone. During our experimentation, we found that because RDRAND is a significant bottleneck in the benchmark experimental loop, results can appear equal for two CSet data structures that otherwise behave very differently. 
%Here, experiments that were run using the XOR-SH* PRNG (Figure 2(d)) indicate that one data structure is superior to the other for throughput results.
Although RDRAND is slow, it can be useful as a source of entropy for faster software PRNGs. The idea of periodically reseeding to introduce additional randomness into a PRNG is discussed by Manssen et al.~\cite{Manssen_2012} and Dammertz~\cite{GPU:conf}. RDSEED is an Intel Secure Key instruction that complements RDRAND and is used to generate high quality random seeds for seeding PRNGs~\cite{RDSEED:link}. RDSEED is slower than RDRAND but is recommended to use for reseeding PRNGs.
We tested a hybrid PRNG solution on the XOR-SH* algorithm where RDSEED is used for reseeding at intervals of every 1 million RNs (\textit{XOR-SH*\_RDSEED}). The results in Figure \ref{RDRAND}(b) indicate comparable throughput results to purely software based PRNGs without reseeding. We compared XOR-SH* reseeding with RDSEED to XOR-SH* reseeding with RDRAND (\textit{XOR-SH*\_RDRAND}), and there is a small drop in performance with RDSEED.
\subsection{PRNG Recommendations}
%\textbf{Recommendation:}
\label{PRNGrecommendation}
%Some important points to consider for synthetic microbenchmarks. (1) Use 64b-bit PRNGs over 32-bit; experiments with billions of operations per second are bound to enter cycles with a 32-bit PRNG. (2) Do not exclusively use cryptographic or hardware-based PRNGs alone. These higher quality, slower PRNGs should be used for periodic reseeding to inject randomness back into a simpler PRNG at regular intervals. (3) A pre-generated array of random numbers should be used only when the PRNG is computatively complex and in-place random number generation will slow down the experiment. (4) Experiments that rely on bitwise randomness in random numbers should first assess various bit positions using measures such as the $n^{th}$ bit summation result.

%we believe a hardware RNG can assist in mitigating some of the low quality randomness results seen in common PRNGs and do not cause a heavy execution penalty.
Massively parallel, high throughput experiments require billions of random numbers to be generated per second, which pushes the limits on PRNGs of our time. Some important points to consider for PRNG usage in microbenchmarks: (1) Hardware RNGs provide an external source of entropy, however, they are impractical for use in high speed concurrent microbenchmark experiments as the performance results are greatly impeded by RN generation time. (2) A pre-generated array of RNs is also counterproductive due to penalties associated with cache access. A pre-generated array of RNs may be useful if the PRNG algorithm is complex and in-place RN generation is too expensive. (3) For synthetic microbenchmark experiments we recommend two PRNG instances per thread; one for generating random values during the experiment and one for injecting new entropy into the first PRNG (periodic reseeding). If periodic reseeding is used every 1 million keys, there is a low, intangible impact on performance. If RDSEED or RDRAND are not available, we recommend using a high quality cryptographic PRNG for the $2^{nd}$ PRNG. (4) Experiments that rely on bitwise randomness in RNs should first examine the set of generated numbers for periodic behaviour across various bit positions. (5) Last, in an era where data structures are performing billions of operations per second, we also think it's important to use PRNGs with at least 64 bits of state to avoid repeating the same sequence of generated keys on a time scale of seconds. 
% The XOR-SH* PRNG is fast and versatile with long period suitable for concurrent microbenchmark tests with key ranges into the millions or even billions of keys. We recommend a software PRNG, such as the XOR-SH* PRNG, with periodic reseeding with a hardware RNG such as RDRAND.
%\textbf{Recommendation}: 
%Although the FNV1a odd-even patterns could be adjusted with a quick work-around, such as employing two different PRNG objects per thread in Setbench, we sought to develop a better understanding of PRNG \textit{quality}, \textit{performance} and overall \textit{suitability} for usage in concurrent synthetic microbenchmarks. In this work we compare FNV1a to other commonly employed software PRNGs, Murmurhash3 (MM3), Murmurhash2 (MM2), XOR-SH*~\cite{XOR-SH*}, MT (Mersenne Twister)~\cite{mersennetwister}, and MRG~\cite{mrgprng}.

%%**UPDATED HERE **
\section{Towards Better Microbenchmarks} \label{other_microbenchmark_design}

In Section \ref{microbenchmark_design}, \ref{APP:BST-TK} and \ref{PRNGrecommendation} we gave some recommendations for good benchmark design that were informed by our study of Ascylib, Setbench and Synchrobench. Setbench was designed with many of those recommendations in mind, and underwent relatively few changes as a result of our study. In this section, we give some additional recommendations and highlight features of Setbench that promote high quality experiments.
% discussed microbenchmark design considerations induced by experiments in this work.
% We now focus on additional microbenchmark design issues and provide some recommendations regarding optimal microbenchmark features. Detailed characteristics of the  Setbench microbenchmark that are centered around good microbenchmark design are discussed in Section \ref{setbench_detail}.

%Further details the Setbench microbenchmark that are centered
%around good microbenchmark design are discussed in greater detail in Section

%eatures in the Setbench microbenchmark that are centered
%around good microbenchmark design are discussed in greater detail in Secti

\subsection{Additional Recommendations} \label{other_microbenchmark_design_details}
% As the research area of concurrent data structures continues to expand
% rapidly, the microbenchmarks used to test these data structures must equally evolve to support the increasing demands. The need for robustness and accuracy in concurrent data structure microbenchmarks is more important than ever. We have the following additional recommendations regarding microbenchmark design:

\paragraph*{More expressive ADTs}
\vspace{-2.0mm}
Today, many CSet data structures support range query operations and other interesting operations such as clone and size, and benchmarks should consider including support for these operations.
It is not necessary for every operation to be implemented by every data structure, but providing a framework for additional operations to be included in experiments may encourage research in this direction.

Similarly, we encourage support for maps (also called dictionaries), which associate a value with each key, and support for large and/or variable-sized keys and values.
This could encourage evaluations that span data structures published in distributed computing venues and those published in database and data management venues (e.g.,~\cite{arulraj2018bztree,leis2013adaptive,levandoski2013bw,mao2012cache})---evaluations that are desperately needed in our opinion.
\vspace{-2.5mm}
\paragraph*{Starvation-aware experiments}
\vspace{-2.0mm}
Note, however, that some care is needed in experimenting with range queries, and any other types of long running operations that are prone to starvation.
Consider a workload where threads perform, say, 49\% insert, 49\% delete and 2\% range queries spanning the entire range of keys contained in the data structure.
One would expect such range queries to be starved by updates, but in practice we find they are not!
The trick is that \textit{each} thread will perform only so many updates before performing a range query.
So, if all range queries are perpetually starved, while updates succeed, eventually \textit{all} threads will be executing range queries, and they will all succeed in a batch.
This behaviour makes starvation seem like less of a problem than it might be in the real world, where there might \textit{never} be a time when there are no updates in progress.
Experiments involving starvation prone operations should \textit{expose} the effects of starvation, possibly by allowing groups of threads to be dedicated to starvation-prone and non-starvation-prone operations respectively (see, e.g.,~\cite{arbel2018range}).
\vspace{-2.0mm}
\paragraph*{Pinning threads}
\vspace{-2.0mm}
%and the ability to specify update rates \textit{unevenly} across insert and delete operations. (1) Microbenchmark design should incorporate testing of range queries operations and allow insertion and deletion rates to be specified separately.
%In addition to an update and retrieval workload, microbenchmarks should provide support for range query operations occurring in parallel by multiple threads. A range query operation allows threads to provide a start and end range and a list is returned of what keys within the range are present in the data structure.
In Section \ref{Ascy_threads}, we disabled thread pinning in all microbenchmarks in order to have consistency across all experiments.
We recommend pinning threads to improve consistency of experiments, so for example, when you run 48 threads on a system with four 48-thread sockets, your threads run on a predictable set of cores, rather than, e.g., being clustered on one socket in one execution, and spread across three sockets in another.
Additionally, thread pinning should be used to clearly expose the performance impact of hyper threading and the effects on non-uniform memory architectures in performance graphs.
Thread pinning in benchmarks has been discussed in more detail by Gramoli~et~al.~\cite{synchro,threadpinning_Gramoli} and Brown~\cite{rare}.
\vspace{-2.0mm}
\paragraph*{Non-uniform key distributions}
\vspace{-2.0mm}
Benchmarks should also consider incorporating various distribution generators for keys and values, rather than limiting experiments to uniform randomness.
Researchers should consider using Zipfian, binomial, exponential or other skewed distributions in their experiments~\cite{NAIR2018107}.
Distribution generators should be implemented efficiently, and sanity checks should be performed to ensure that the rate of key/value generation is not a bottleneck.
\vspace{-2.0mm}
\paragraph*{Uniform memory reclamation}
\vspace{-2.0mm}
Research in safe memory reclamation for CSets has consistently demonstrated that CSet performance can depend heavily on the algorithm for reclaiming memory (see, e.g.,~\cite{citeSB1,nikolaev2021brief,sheffi2021vbr,singh2021nbr,wen2018interval}).
For this reason, memory should be reclaimed similarly across all data structures evaluated.
In some cases, ad-hoc memory reclamation is tightly integrated in a CSet, but benchmarks should offer a fast, easy-to-use memory reclamation algorithm to promote uniformity wherever possible.
%Microbenchmarks should attempt to ensure memory reclamation algorithms are similar across all data structures in order to avoid biased performance results due to differences in memory reclamation implementations.
\vspace{-2.0mm}
\paragraph*{Performance tools}
\vspace{-2.0mm}
Benchmarks should make a best effort attempt to \textit{automatically} gather lightweight systems-level performance data, such as cache misses per operation, total cycles per operation, and peak memory usage.
We suggest incorporating a library for performance monitoring such as the Performance Application Programming Interface (PAPI)~\cite{papi_article}.
We think it is crucial that these measurements are not only automatically gathered, but automatically \textit{visualized}.
Ideally, graphs for CSet throughput results and for systems level performance monitoring would be produced by default, at the same time, and would be visible in the same place.
``Easy to check'' is good.
``Difficult to ignore'' is better.

\subsection{Benchmarking Advances in Setbench} \label{setbench_detail}
Setbench was specifically designed to address all of the recommendations above, featuring a new 256-bit PRNG, range query support (with support for independent range query threads and update threads), the ability to specify thread pinning policies at the command line, fast Zipfian and Uniform key distributions, uniform epoch based memory reclamation, and integration with a rudimentary implementation~\cite{yu2015evaluation} of TPC-C and YCSB application benchmarks.
It also includes a large collection of powerful tools for debugging, running experiments and analyzing performance, as well as automatic containerization for artifact evaluation.
\vspace{-2.0mm}
\paragraph*{Collecting user defined statistics}
\vspace{-1.0mm}
Debugging and performance analysis are extremely time consuming, and often researchers are limited in how much investigation they can do by the time it takes to modify their code to record specific events in their data structure.
These events can be quite varied.

For example, one might want to answer a simple question like: in a lock-free algorithm, \textit{how often do threads help complete other threads' operations?}
Or, in an algorithm that uses epoch based memory reclamation, where objects are reclaimed in batches, one might want to answer a much more complex question---\textit{how to produce a logarithmic histogram showing the distribution of the sizes of the first 10,000 batches reclaimed by each thread in an execution.}
Setbench's global stats library (\textbf{gstats}) makes it fast and easy to explore such questions.
        
To emphasize how easy gstats makes this, to implement the latter, one would first create a gstats statistic that is accessible globally (throughout all files in the entire benchmark), by adding the following code to a file in Setbench called \texttt{define\_global\_statistics.h}:
\vspace{-1.0mm}
\begin{verbatim}
    gstats_handle_stat(LONG_LONG, epoch_batch_size, 10000, \
        { gstats_output_item(PRINT_HISTOGRAM_LOG, NONE, FULL_DATA) })  \
\end{verbatim}        
\vspace{-1.0mm}
\noindent In essence, this efficiently allocates global per-thread arrays of 10,000 elements, and specifies that their contents should be used to build a logarithmic histogram. Whenever a thread \texttt{T} reclaims a batch of size \texttt{n}, it can append the batch size \texttt{n} to its array by invoking:
\vspace{-1.0mm}        
\begin{verbatim}
    GSTATS_APPEND(T, epoch_batch_size, n);
\end{verbatim}
\vspace{-1.0mm}
\noindent These simple modifications result in the following new output when the benchmark is run:
\vspace{-1.0mm}
\begin{small}
\begin{verbatim}
    log_histogram_of_none_epoch_batch_size_full_data=[...]
            [2^00, 2^01]: 71905
            (2^01, 2^02]: 206257
            (2^02, 2^03]: 307829
            (2^03, 2^04]: 469972
            [...] // output truncated to save space
\end{verbatim}
\end{small}

\noindent Furthermore, scripts are included to plot bar graphs and line graphs from any data collected with gstats.
In this case, assuming the output above is in \texttt{data.txt}, one would simply run: \texttt{trial\_to\_plot.sh data.txt epoch\_batch\_size}, which would create a PNG file.
\vspace{-2.0mm}
\paragraph*{Running Experiments and Plotting Results}
\vspace{-1.0mm}
Setbench also offers a powerful suite of Python scripts for running experiments and automatically plotting their results.
Example run scripts that are suitable for CSet research are included.
They produce MatPlotLib graphs of throughput \textit{and many systems level performance metrics}, such as L3 cache misses per CSet operation, cycles per operation, and peak memory usage.
Scripts are also available for several papers published by our group.
The development of these scripts focused on conciseness, expressiveness and flexibility, and the scripts could be adapted to drive completely different benchmarks in different domains.

At a high level, to use these scripts, one defines a sequence of experimental \textit{parameters}, and for each parameter, one specifies a list of values the parameter should take on. % a sequence of experimental parameters, and a list of values for each parameter. % each should take on.
One then specifies a \textit{run command} for the benchmark, and specifies how the parameters should be supplied to the run command.
The command is run for each combination of parameters, and the output of each run is stored in an individual file.
The scripts then process each file, and extract lines of the form ``NAME=DATA'' to produce columns in a \textbf{sqlite database}.
As part of this process, data is \textit{validated} according to user specified rules such as \texttt{(`total\_throughput', is\_positive)} or \texttt{(`validate\_result', is\_equal(`success'))}.
Failed validation causes (colourful!) warnings to be emitted, and warnings can also be queried later from the sqlite database.

The scripts expose functions for easily producing plots (bars, lines, histograms and heatmaps) from the sqlite database simply by specifying which columns of data should be used for the x-axis and y-axis.
Additional columns can be specified and graphs will be produced for every combination of values in these columns.
Filters can also be specified to add to the SQL WHERE clauses in the queries that underpin plot generation.

In short, a single command \texttt{run\_experiment.py [your\_experiment.py]}, depending on its arguments, can compile (\texttt{-c}), run (\texttt{-r}), create the sqlite database (\texttt{-d}), produce graphs (\texttt{-g}) and create an HTML website (\texttt{-w}) organizing them into sections for convenient viewing.
Clicking a graph on the website drills down to the rows of data the graph was built from, and clicking a row shows the raw text output for that run.
A generated example website can be viewed at: \url{https://cs.uwaterloo.ca/~t35brown/setbench_example_www}.
Results in the sqlite database can also be queried conveniently from the command line using SQL (e.g., \texttt{run\_experiment.py your\_experiment.py -q "select * from data"}).
A wide range of additional capabilities are documented in extensive Jupyter notebook tutorials. % demonstrating capabilities of these python scripts.

\section{Conclusions}\label{conclusions}
We hope this work encourages further research into how best to design benchmarks for concurrent data structures. Setbench was carefully designed to mitigate many of the problems we are aware of, but there are surely more benchmarking pitfalls yet to be discovered in this area. We also encourage researchers to try using Setbench for their own experiments, because its features make it much easier to drill down to the root causes of performance anomalies. After designing an algorithm, proving correctness, and implementing it, there is often little time left to do systems level performance analysis. Our hope is that by improving tools and automating the collection and graphing of key performance metrics, we can improve the quality of experiments without unduly burdening researchers in this area. 
\newpage

%%
%% Bibliography
%

%% Please use bibtex, 

%\bibliography{may2022_microbenchmarks}
\bibliography{Final_arxiv}

\newpage

%We may have long periods within the experiment where all the random numbers strongly favour bit 3 being %equal to 0.
%or periods where the keys are primiarly positve... and then periods where the keys are negative.
\begin{appendices}

%\appendix

%%%%%%%%%%%%%%%%%%%%%%%%%%%%%%%%%%%%%%%%%%%%%%%%%%%
\begin{figure}
    %\centering
    \hspace{-0.9in}
    \begin{tabular}{cccc}
     (a) Max Resident & (b) Throughput & (c) Max Resident & (d) Throughput \\

    %\rotatebox{90}{\textbf{10   10   10  10}}
    \hspace{-4mm}
    \includegraphics[width=\imgwidth]{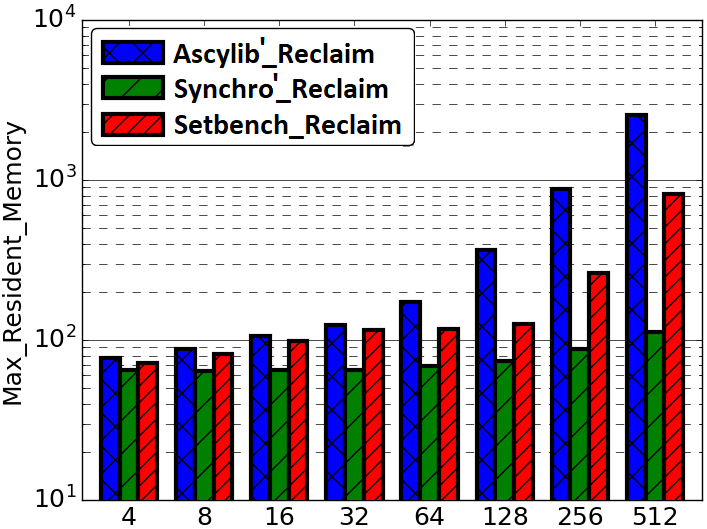} & 
    \hspace{-4mm}
    \includegraphics[width=\imgwidth]{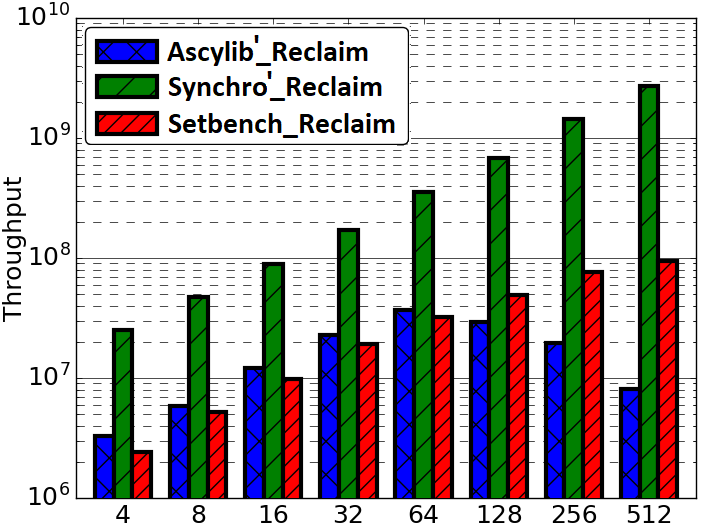} &
    \hspace{-4mm}
    \includegraphics[width=\imgwidth]{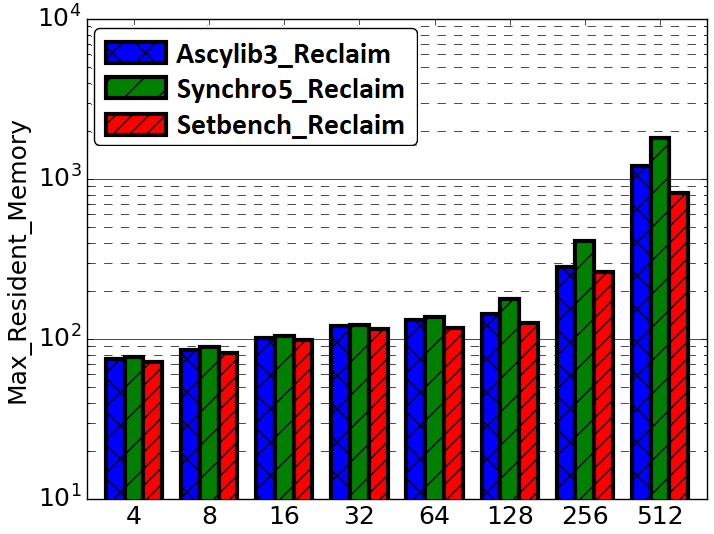} &
    \hspace{-4mm}
    \includegraphics[width=\imgwidth]{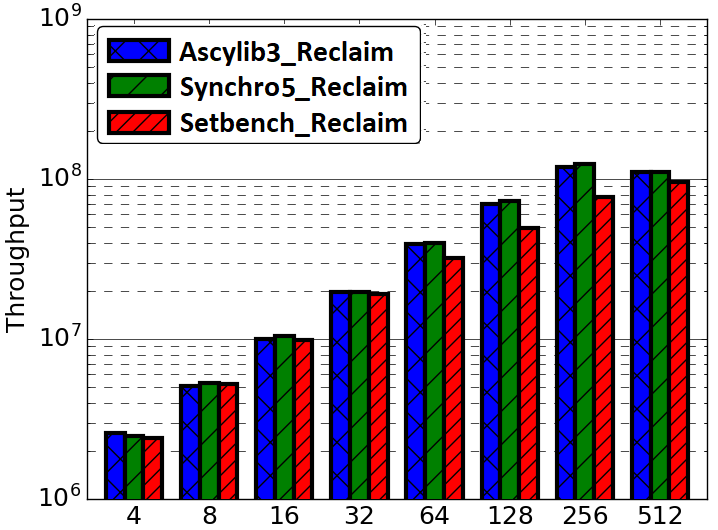} \\

    \end{tabular}
    \caption{Maximum Resident Memory and Throughput results on a 2 million key range at 100\% update rate when the lock-free BST implementation is common across all experiments. Figures (a) and (b) show results without any microbenchmark modifications. Figures (c) and (d) include all microbenchmark modifications to Synchrobench and Ascylib as discussed in Sections 4.2 and 4.3. }
   \label{reclaim_throughput}
\end{figure}
%%%%%%%%%%%%%%%%%%%%%%%%%%%%%%%%%%%%%%%%%%%%%%%%%%%
\section{Lock-Free BST}  \label{appendixA}
In Section \ref{memrec}, we investigated the total memory usage of the Ascylib and Setbench microbenchmarks when testing the Lock-free BST. To better understand the role of each microbenchmark algorithm on maximum resident memory results, we test the microbenchmarks when the lock-free BST data structure implementation is equalized and memory reclamation settings are turned on. The algorithm for epoch-based memory reclamation is also identical in each microbenchmark.
Figures \ref{reclaim_throughput}(a) and (b) test Synchrobench and Ascylib without any microbenchmark modifications but include only an equalized data structure implementation (\texttt{Synchro'} and \texttt{Ascylib'} with reclamation). Ascylib throughput results decline after a thread count of 64. Synchrobench maximum resident memory values are much lower due to missing insert operations in the test loop. Figures \ref{reclaim_throughput}(c) and (d) include all microbenchmark modifications to Synchrobench and Ascylib as implemented in \texttt{Ascylib3} and \texttt{Synchro4}. We see performance values across the microbenchmarks are more consistent when we account for microbenchmark differences. This illustrates the significant impact of microbenchmark idiosyncrasies on both throughput results and memory reclamation. \\
%%OPODIS move OUT OF APPENDIX
%\label{reclaim_lockfree}

%We first compare the memory reclamation implementations in Setbench and Ascylib by setting Ascylib's GC setting to true, and Setbench epoch based reclamation is turned on. Figure \ref{reclaim}(a) illustrates differences in each microbenchmark's ability to reclaim memory as the thread count increases and cores are oversubscribed. Ascylib's memory usage surpasses that of Setbench by over one order of magnitude, particularly as the thread count increases. Throughput results are relatively equal, however, the high maximum resident memory values may render Ascylib experiments infeasible in some settings. \\

\section{PRNGs in Microbenchmark Experiments}
In Section \ref{randomsection}, we discussed optimal PRNG usage during high-speed microbenchmark experiments. In this section we perform additional PRNG experiments for assessing methods for high quality random number (RN) generation. We introduce a tool for assessing bitwise patterns that may exist in the sequence of RNs generated by PRNGs. We further investigate the benefit of periodically reseeding a software PRNG with a high quality RNG as a means of providing new entropy.

\subsection{Bitwise PRNG Analysis}
\label{appendixB}
%The original problem discovered with the FNV1a algorithm (even-odd numbers) can be analogized to a pattern where the lowest order bit of a 64-bit number is alternating between on and off. FNV1a returns a series of numbers where the lowest bit strictly switches from 0 to 1, or from 1 to 0 at each call to \texttt{next}. 
As we saw in Section \ref{fnv1a_anomaly}, FNV1a generates numbers that alternate in the least significant bit generating an odd-even pattern of numbers. To test if similar behaviour occurs in other bit positions, or if other PRNGs suffer from similar problems, we created a tool to visualize patterns in the individual bits of randomly generated numbers.
We repeatedly generate RNs and plot a curve of results representing bitwise variance for each bit of the RN. Specifically, for each bit position \texttt{n}, we maintain a running sum that is incremented when the \texttt{$n^{th}$} bit of a generated RN is set (equal to 1), and decremented when the \texttt{$n^{th}$} bit is off. That is, for an \texttt{$n^{th}$} bit equal to 1, the sum is incremented by +1. If the \texttt{$n^{th}$} bit is equal to 0, the sum is incremented by -1. Therefore, a positive sum indicates a greater number of RNs had the \texttt{$n^{th}$} bit equal to 1, a negative sum indicates a greater number of RNs had the \texttt{$n^{th}$} bit equal to 0. An \texttt{$n^{th}$} bit sum converging to zero indicates that the bit was evenly assigned to 1 and 0 throughout the set of generated RNs.
We will illustrate with an example: If the \texttt{$n^{th}$} bit is the lowest order bit (bit zero), and we generate 1000 RNs and obtain a final \texttt{0-bit} sum of 100, this indicates that there were 100 more random numbers with the zero bit equal to 1 than equal to 0.

We compare \texttt{$n^{th}$} bit summation results across all bit positions of 32-bit and 64-bit RNs as a simple measure of randomness. It is a measurable indicator of a PRNG favouring RNs with certain bit position set or not set. Figure \ref{fig:suml} illustrates \texttt{$n^{th}$} bit summation results from generating 1000 random numbers using FNV1a, XOR-SH*, and MM2 for the first eight least significant bits. We experimented also with up to 1 billion generated RNs and observed similar results, however, the results are difficult to visualize. We display results only for the first 1000 RNs.
%We created a tool to assess bitwise occurrence patterns of random numbers generated by a PRNG. We keep a running \texttt{sum} of the $n^{th}$ bit of random number based on this bit being on or off. The \texttt{$n^{th}$ bit sum} is incremented a -1 if the $n^{th}$ bit is equal to 0 and a +1 if the $n^{th}$ bit is equal to 1. For example, in the case of FNV1a, we assign a 1 or -1 to every random number based on the lowest bit; the sum of 1's and -1's is computed to give a measurable indicator of a PRNG favouring odd or even n
\iffalse
\begin{figure}
    \centering
    \includegraphics[width=\linewidth]{figure/Rdrand_4_fig.png}
    \caption{RDRAND Performance results. Hardware vs Software PRNG (a) RDRAND throughput compared to software PRNGs. (b) Reseeding XOR-SH* PRNG with RDRAND or RDSEED every 1 million random numbers indicates no strong penalty for reseeding with a hardware RNG. }
    \label{RDRAND}
\end{figure}
\fi
As is expected, FNV1a tested on a single thread produces an overall \texttt{0-bit} sum of 0 since there are exactly an equal number of even and odd random numbers generated (assuming the total number of random numbers is a multiple of 2). The FNV1a case is an exception. If a PRNG produced an equal or an almost equal number of odd and even numbers, this is not necessarily an undesirable quality. The problem with FNV1a is the ordered pattern of odd-even RN generation. This is undesirable in any PRNG.
In addition to an alternating pattern in the 0-bit, the FNV1a algorithm indicates patterns within higher order bit positions as illustrated in Figure \ref{fig:suml}(a), where FNV1a is compared to other PRNGs across bit positions 0 to 7. FNV1a indicates clear periodic behaviour of other bit positions equaling 0 or 1 overtime. As the bit position increases, the length of the summation cycle gets longer, but a pattern still exists. The XOR-SH* algorithm indicates no strong summation patterns in the 8 represented bits, however we can see that numbers generated by the XOR-SH* PRNG slightly favour the $7^{th}$ bit equaling a 1 since the final sum ends at close to 50. $N^{th}$ bit patterns can play an important role in experiments where bit positions are expected to be random and used to make decisions, such as using \texttt{$n^{th}$} bit positions to decide which level of an n-level skip list to perform an insert operation. Bit position patterns in RNs may cause subtle changes in performance results as we have seen with FNV1a.

%%%%%%%%%%%%%%%%%%%%%%%%%%%%%%%%%%%%%%%%%%%%%%%%%%%%%%%%%%%%%%%%
\begin{figure}
    \centering
    
    \includegraphics[width=\linewidth,height=1.7in]{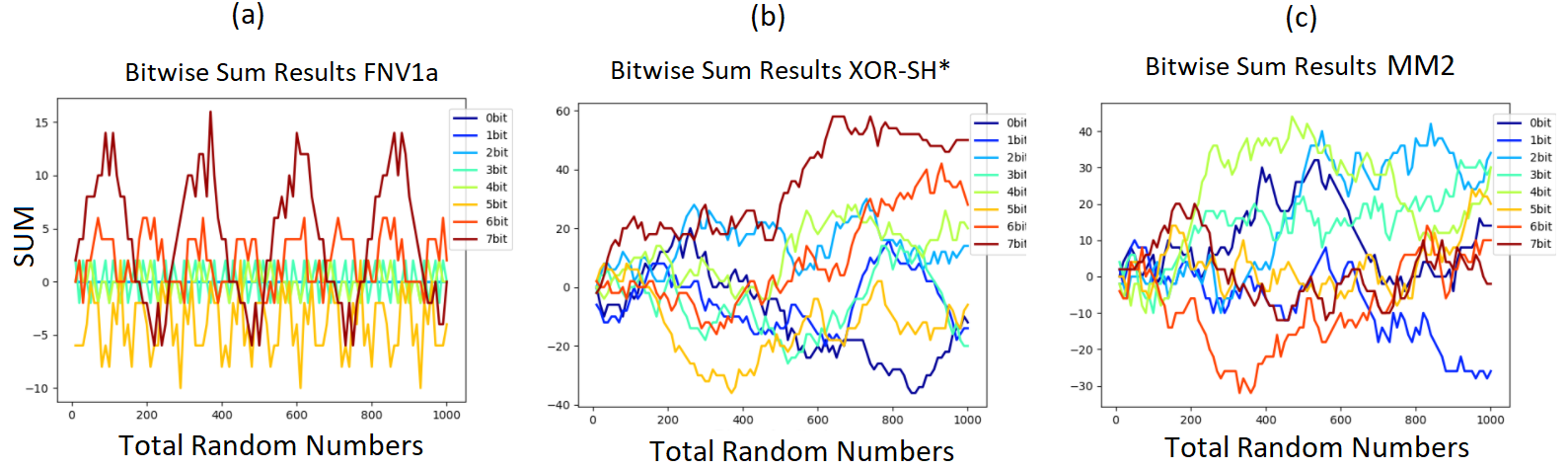}
    \caption{Bitwise Summation Results for the first byte of random numbers generated through PRNGs (a) FNV1a (b) XOR-SH* (c) MM2.}
    \label{fig:suml}
\end{figure}
%%%%%%%%%%%%%%%%%%%%%%%%%%%%%%%%%%%%%%%%%%%%%%%%%%%%%%%%%%%%%%%

\begin{figure}
    \centering
    \includegraphics[width=\linewidth, height=2.4in]{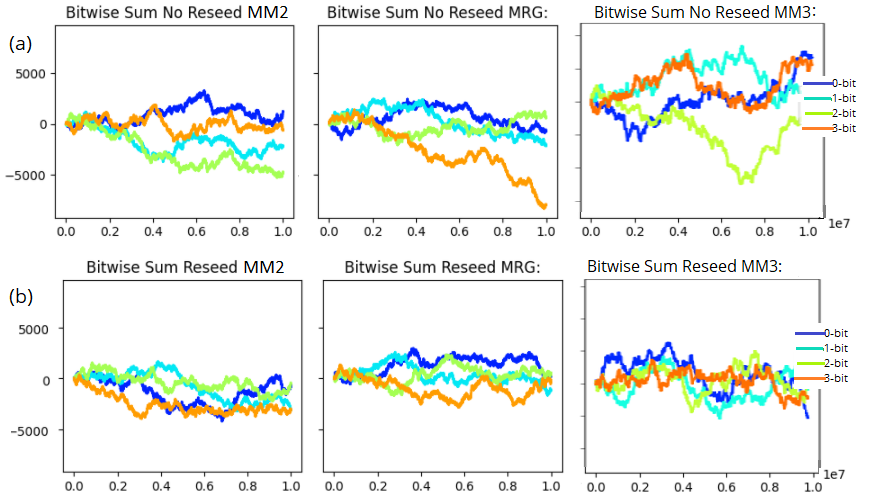}
  \caption{Summation Results for 10 million random numbers: (a) MM2, MRG and MM3 tested without reseeding, (b) MM2, MRG, MM3 tested with reseeding. We can see that using periodic reseeding generates tighter boundaries on the summation results of the first 4 bits of a random number.}
    \label{fig:seedingvsnoseeding}
\end{figure}

We tested summation results across every bit position in 32-bit or 64-bit PRNGs.
It interesting to see the patterns and periodic behaviours that occur overtime in Figures \ref{fig:suml} and \ref{fig:seedingvsnoseeding}. Summation results across the various algorithms can depict negative or positive values over time. This indicates that a PRNG is favouring numbers where the $n^{th}$ bit turned on, in the positive sum case, or favouring numbers with the $n^{th}$ bit turned off, in the negative sum case. 
There may be a tendency towards positive or negative sums during the experiment, however, we consider a PRNG to be \textit{balanced} in the $n^{th}$ bit if summation results converge back towards zero by the end of the experiment.
FNV1a may be completely \textit{balanced} for the \texttt{0-bit} sum, however, this result needs to be considered in context with the odd-even patterns generated by this algorithm.

%%%%%%%%%%%%%%%% end figure %%%%%%%%%%%%%%%%%%% 
%%%%%%%%%%%%%%%% end figure %%%%%%%%%%%%%%%%%%% 
\subsection{Periodic Reseeding}
As discussed in~\cite{GPU:conf}, periodic reseeding can be used to inject entropy into the PRNG at regular intervals. The idea is to provide the PRNGs with a new source of randomness using RDSEED. Our goal was to test if periodic reseeding of software PRNGs led to more stable summation results. We look at comparative results in Figure \ref{fig:seedingvsnoseeding} between three PRNGs tested with and without reseeding on a sequence of 10 million RNs. We see that bitwise summations that tend to trail off in either direction can be mitigated with periodic reseeding. We are looking for summation values that are very high or very low to become more balanced using reseeding. In Figure \ref{fig:seedingvsnoseeding}(a) we see summation values tend to trail off in either the positive or negative direction. This indicates bits are more consistently on or off and there is an imbalance in the expression of the bit. For example, in Figure \ref{fig:seedingvsnoseeding}(a), the MRG PRNG illustrates that bit 3 tends to be off at much higher rates than it is on. Figure \ref{fig:seedingvsnoseeding}(b) illustrate the impact of reseeding on the bitwise summation results. When reseeding in used every 1 million RNs, the PRNGs tend to have a more balanced sequences of RNs with summation results that do not trail off in either direction. We also found in Section \ref{hardware}, that periodic reseeding every 1 million RNs does not have a significant impact on performance.

Researchers employ different PRNGs in microbenchmark experimentation for a variety of reasons. Rather than disposing of one PRNG model for another, we found that periodic reseeding with RDSEED can be used to improve the randomness quality of a PRNG without significantly hindering performance. 
As stated in our conclusions, if RDSEED or RDRAND are not available, we 
recommend using a high quality cryptographic PRNG for the $2^{nd}$ PRNG. \\

\end{appendices}
\end{document}